\documentclass{aa}  

\usepackage[normalem]{ulem} 
\usepackage{threeparttable} 
\usepackage{graphicx}
\usepackage{txfonts}

\usepackage{color}

\begin{document}

   \title{Modeling dust emission in the Magellanic Clouds with \textit{Spitzer} and \textit{Herschel}}

   \author{J\'er\'emy Chastenet
          \inst{1,2}
          \and
          Caroline Bot\inst{1}
          \and
          Karl D.\ Gordon\inst{2,3}
          \and 
          Marco Bocchio\inst{4}
          \and 
          Julia Roman-Duval\inst{2}
          \and
          Anthony P. Jones\inst{4}
          \and
          Nathalie Ysard\inst{4}
          }

   \institute{Observatoire astronomique de Strasbourg, Université de Strasbourg, CNRS, UMR 7550, 11 rue de l’Université, F-67000 Strasbourg, France\\
              \email{jeremy.chastenet@astro.unistra.fr}
        \and
            Space Telescope Science Institute, 3700 San Martin Drive, Baltimore, MD, 21218, USA
        \and
            Sterrenkundig Observatorium, Universiteit Gent, Gent, Belgium
        \and
            Institut d'Astrophysique Spatiale, CNRS, Univ. Paris-Sud, Université Paris-Saclay, Bât. 121, 91405 Orsay cedex, France
        }

   \date{Received...; accepted...}

 
  \abstract
   {Dust modeling is crucial to infer dust properties and budget for galaxy studies. However, there are systematic disparities between dust grain models that result in corresponding systematic differences in the inferred dust properties of galaxies.  Quantifying these systematics requires a consistent fitting analysis.}
   {We compare the output dust parameters and assess the differences between two dust grain models, namely those built by \citet{Compiegne11}, and THEMIS \citep{Jones13, Kohler15}. In this study, we use a single fitting method applied to all the models to extract a coherent and unique statistical analysis.}
   {We fit the models to the dust emission seen by \textit{Spitzer} and \textit{Herschel} in the Small and Large Magellanic Clouds (SMC and LMC). The observations cover the infrared (IR) spectrum from a few microns to the sub-millimeter range. For each fitted pixel, we calculate the full n-D likelihood, based on the method described in \citet{Gordon14}. The free parameters are both environmental ($U$, the interstellar radiation field strength; $\alpha_\mathrm{ISRF}$, power-law coefficient for a multi-U environment; $\Omega^*$, the starlight strength) and intrinsic to the model ($Y_\mathrm{i}$: abundances of the grain species $i$; $\alpha_\mathrm{sCM20}$, coefficient in the small carbon grain size distribution).}
   {Fractional residuals of 5 different sets of parameters show that fitting THEMIS brings a more accurate reproduction of the observations than the \citet{Compiegne11} model. However, independent variations of the dust species show strong model-dependencies. We find that the abundance of silicates can only be constrained to an upper-limit and the silicate/carbon ratio is different than that seen in our Galaxy. In the LMC, our fits result in dust masses slightly lower than those found in literature , by a factor lower than 2. In the SMC, we find dust masses in agreement with previous studies.}
   {}

   \keywords{Infrared: galaxies - Galaxies: ISM - ISM: dust - Magellanic Clouds }

   \authorrunning{J. Chastenet}
   \maketitle

\section{Introduction}
Dust plays a fundamental role in the evolution of a galaxy. For example, it has a large impact on the thermodynamics and chemistry processes, by catalyzing molecular gas formation. Dust grains are the formation sites of molecular hydrogen (H$_2$). Recent studies \citep{LeBourlot12, Bron14} show that the efficiency of H$_2$ growth is sensitive to dust temperature and grain size distribution. 

Dust can also be used a gas tracer in nearby and distant/high-z galaxies with knowledge of the gas-to-dust ratio (GDR). The dust abundance reflects the chemical history of galaxies, and so it is important that we measure how it varies with environment. Various studies have computed global gas and dust masses in galaxies, finding a clear trend between the GDR and metallicity \citep[e.g.][]{Engelbracht08, Engelbracht08Err, Remy-Ruyer14}.  Such studies fit the observed IR emission with dust grain models to measure dust masses.  Studies of the most nearby galaxies offer the possibility to estimate these quantities at high resolution on a pixel-by-pixel basis. For example, \citet{RomanDuval14} investigated the spatial variations of the GDR in the Magellanic Clouds, in particular its evolution from diffuse to dense phase. 
To comprehend the dust impact on other processes and features in the ISM, it is of crucial importance to understand its physical state and composition, including minimal and maximal grain sizes, using, for example, dust grain models.

Full dust grain models are numerous \citep[e.g.][]{Oort46,MRN77,DBP90,Clayton03, Zubko04,DraineLi07,Compiegne11,Galliano11,Jones13,Kohler14,Kohler15} and vary from one another by the definition of dust composition, size distribution of grains, laboratory-based data for optical properties, and are not necessarily constrained by the same observational references. 
Development of such models are of crucial importance since dust strongly affects the radiative transfer of energy in a galaxy. To fathom the emerging spectral energy distribution (SED) from X-Rays to the IR, we need to understand dust properties and their variations, and accurately model them.
Many studies attempt to fit observational data with a given model and fitting technique. To this day, the widely accepted description of dust involves two main chemical entities : carbonaceous grains, which usually show both amorphous and aromatic structures, and silicate grains, with metallic-element inclusions to agree with the observed composition.

Over the past decades, observational knowledge of dust have tantalizingly increased. In the infrared, the first all-sky survey provided by the \textit{Infrared Astronomical Satellite} (\textit{IRAS}) and the \textit{Cosmic Background Explorer} (\textit{COBE}) were major breakthroughs. The \textit{Spitzer Space Telescope} and the \textit{Herschel Space Observatory} gave a tremendous amount of data constraining the emission from dust at a higher resolution.
In the ultraviolet, continued observations and analysis of extinction \citep{Cardelli88, Cardelli89, Mathis90, Fitzpatrick05, Cartledge05, Gordon03, Gordon09} and depletions \citep{Jenkins09, Tchernyshyov15} have shown that large variations in dust properties exists from one line of sight to the next, and between galaxies. 

Two of the most nearby galaxies within the Local Group, the Small Magellanic Cloud (SMC) and the Large Magellanic Cloud (LMC; together, MCs) have been extensively studied in many surveys. At respectively 62~kpc \citep{Graczyk14} and 50~kpc \citep{Walker12}, they span a range of properties which make them good targets to study different environments (e.g. different stellar populations). In particular, their metallicities are lower than our galaxy (SMC: 1/5~$\mathrm{Z}_\odot$; LMC: 1/2~$\mathrm{Z}_\odot$ \citep{RusselDopita92}), and respectively lower and higher than the threshold of $\sim~1/3-1/4~Z_\odot$  that marks a significant change in the ISM properties \citep{Draine07}. The improvement of spatial and ground-based instruments have pushed the limit of our knowledge of these galaxies, allowing us to test more sophisticated theories about ISM evolution.

Observations show that the infrared SEDs of the MCs differ from those seen in the Milky Way (MW). At (sub-)millimeter and centimeter wavelengths, dust is well modeled by a blackbody spectrum modified by a power-law. Many investigations have identified this trend by pointing out ``excess'' emission in the far-infrared (FIR) to centimetric \citep[e.g.][]{Galliano03, Galliano05, Bot10b, Israel10, Gordon10, Galliano11, Gordon14}. In those models, it means that the spectral emissivity index $\beta$ is lower in the MCs than in the MW. \citet{Reach95} suggested this excess in the MW comes from cold dust, but rejected this hypothesis as the dust mass needed to account for such an emission (with dust at very low temperature) would be too high to be realistic, and violate elemental abundances. The current theory points towards different power-law (i.e. different spectral indices) in the expression of the emissivity, in different wavelength ranges (e.g. a `broken-emissivity' modified blackbody model).
Another excess has been identified at $70~\mu$m, with respect to the expected emission from MW-based dust models. The works of \citet{Bot04} and \citet{Bernard08} linked this excess to a different size distribution and abundance of the very small grains whose emission is dominant at these wavelengths. 
The infrared peak ($100~\mathrm{\mu m} \leqslant \lambda  \leqslant 250~\mathrm{\mu m} $) also varies between the MW and the MCs and tends to be localized at shorter wavelengths in the SMC. This tendency may be due to the harder radiation fields in the SMC, once again suggesting that the models based on MW observations may not fit the SMC dust emission.  

Although we may identified common behaviour with different models (e.g. sub-millimeter excess), the same models do not agree on all properties (e.g. dust masses). It is difficult to determine whether the differences in dust studies arise from the intrinsic descriptions of the dust models, or the statistical treatment of the fitting algorithm, or both.
In this paper, we use current dust grain models to fit the MIR to sub-millimeter observations of the MCs. Our goal is to quantitatively measure the discrepancies between the models used in a common fitting scheme, and assess which part of the SEDs can be reproduced best with a given set of physical inputs. To do so, we base our effort on the work of \citet{Gordon14}. In their study, they focused on fitting three models to the \textit{Herschel} HERITAGE PACS and SPIRE photometric data : the Simple Modified BlackBody, the Broken Emissivity Modified BlackBody and the Two Temperatures Modified BlackBody (SMBB, BEMBB and TTMBB, respectively). They identified a substantial sub-millimeter excess at 500$\mathrm{\mu m}$, likely explained by a change in the emissivity slope. They built grids of spectra, varying parameters for a given model (e.g. for the SMBB model, they allow the dust surface density, the spectral index, and dust temperature to vary). They adopted a Bayesian approach to derive, for each spectrum, the multi-dimensional likelihood assuming a multi-variate Normal/Gaussian distribution for the data to assess the probability that a set of parameters fits the data. The residuals and derived gas-to-dust ratio favor the BEMBB model, which best accounts for the sub-millimeter excess. We use the same statistical approach in this study, although we extend the observational constraints to shorter wavelengths (Section \ref{SecData}). Hence, we must account for smaller dust grains and ``full'' models, and we make use of the DustEM tool \footnote{\url{http://www.ias.u-psud.fr/DUSTEM/}} \citep{Compiegne11} to build our own grid of physical dust models (Section \ref{SecTools}). We then compare the different models used based on residuals characteristics (Section \ref{SecModelComp}) and derive physical properties and interpretation (Sections \ref{SecResults} and \ref{SecDiscussion}).

\section{Data}
\label{SecData}
In this study, we fit the dust emission in the Magellanic Clouds.
The MIR, FIR, and sub-millimeter images used in this study are taken from the \textit{Spitzer} SAGE-SMC \citep[Surveying the Agents of Galaxy Evolution;][]{Gordon11a} and SAGE-LMC \citep{Meixner06} Legacies and the \textit{Herschel} HERITAGE Key Project \citep[The Herschel Inventory of the Agents of Galaxy Evolution;][]{Meixner13, Meixner13err}.
The SAGE observations were taken with \textit{Spitzer Space Telescope} \citep{Werner04} photometry instruments: the Infrared Array Camera \citep[IRAC;][]{Fazio04} provided images at 3.6, 4.5, 5.8 and 8.0~$\mathrm{\mu m}$ and the Multiband Imaging Photometer for \textit{Spitzer} \citep[MIPS;][]{Rieke04} providing images at 24, 70 and 160~$\mathrm{\mu m}$. The observations cover a $\sim {30~\mathrm{deg}}^2$ region for the SMC and $\sim {50~\mathrm{deg}}^2$ for the LMC. 
Data in the FIR to sub-millimeter were taken with PACS \citep[Photoconductor Array Camera and Spectrometer;][]{Poglitsch10} and SPIRE \citep[Spectral and Photometric Imaging Receiver;][]{Griffin10} on-board the \textit{Herschel Space Observatory} \citep{Pilbratt10}, providing images at 100, 160, 250, 350 and 500~$\mathrm{\mu m}$.  The observations cover the same regions as the \textit{Spitzer} data.

For this study, we used the combined \textit{Spitzer} and \textit{Herschel} set of bands to cover the IR spectrum.  The combined bands are from IRAC 3.6, 4.5, 5.8, \& 8.0~$\mathrm{\mu m}$, MIPS 24, \& 70~$\mathrm{\mu m}$, PACS 100 \& 160~$\mathrm{\mu m}$, and SPIRE 250, 350 \& 500~$\mathrm{\mu m}$. Thanks to the custom de-striping techniques used to process the HERITAGE data \citep[see][for details]{Meixner13}, the PACS 100 data combines the resolution of Herschel with the sensitivity of IRAS 100. Similarly, the PACS 160 image was merged with the MIPS 160 image.

Like \citet{Gordon14}, first, all the images were convolved using the \citet{Aniano11} kernels to decrease the spatial resolution of all images to the resolution of the SPIRE 500~$\mathrm{\mu m}$ band of $\sim 36''$. Next, the foreground dust Milky Way dust emission was subtracted. To do so, we built a MW dust foreground map using the MW velocity \ion{H}{I} gas maps from \citet{Stanimirovic00} for the SMC and \citet{StaveleySmith03} for the LMC. To convert the velocity gas maps to a dust emission map, we use the \citet{Compiegne11} model. We derive conversion coefficients from \ion{H}{I} column to MW dust emission, and subtract the resulting maps to the data.

After this processing the PACS observations show a gradient across the images. We removed this gradient by subtracting a two-dimensional surface estimated from background regions in the images.
We chose regions outside of the galaxies (and bright sources) to evaluate a ``background'' plane and then subtract it on all the images. For the LMC, the observations did not extend beyond full disk and this introduces a larger uncertainty in the final background subtracted images. The SMC observations extend beyond the galaxy and we have access to regions on the images fully outside the galaxy.
Finally, we rebin the images to have a pixel scale of $\sim{56}''$ that is larger than the FWHM of the SPIRE 500~$\mathrm{\mu m}$ band to provide nominally independent measurements for later fitting.

\section{Tools and computation}
\label{SecTools}
\subsection{DustEM}
The DustEM tool \citep[][]{Compiegne11} outputs emission and extinction curves calculated from dust grains description (optical and heating properties) and size distributions, and environment information. We use the DustEM IDL wrapper \footnote{available at \url{http://dustemwrap.irap.omp.eu/}} to generate full model grids with a large number of emission spectra. The wrapper forward-models the observations, by multiplying the model SED with transmission curves. We use two dust models in our study, based on the work from \citet{Compiegne11} and \citet{Jones13} updated by \citet{Kohler14}.

\begin{figure*}
    \begin{center}
        \includegraphics[width=0.9\textwidth]{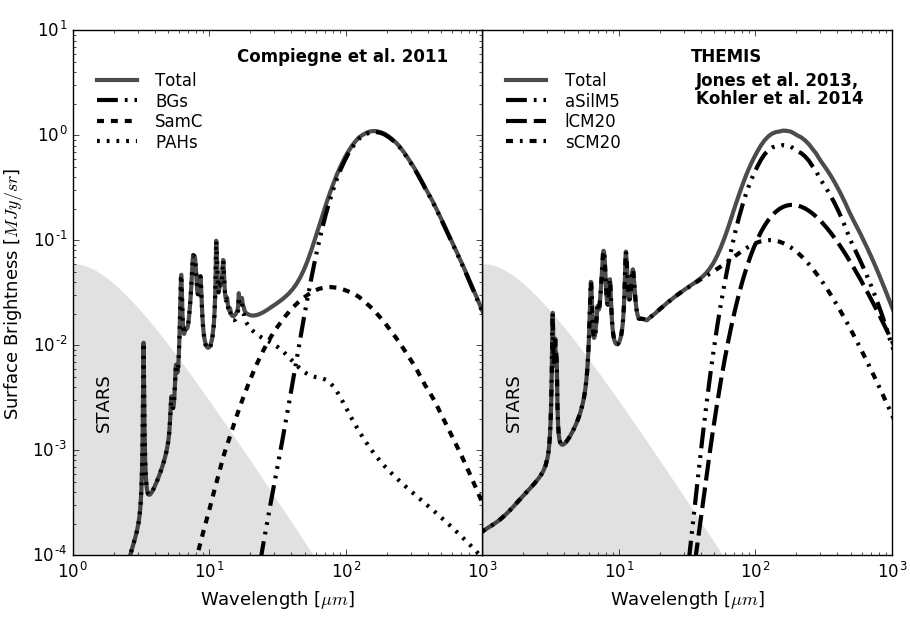}
    \end{center}
    \caption{\small{Original \citet{Compiegne11} (MC11 model) (left) and THEMIS (right), for U=1 (from \citet{Mathis83}) and $N_\mathrm{H}=10^{20}~ \mathrm{H~cm}^{-2}$. They vary by the total number of components and optical and heating properties. The `stars' component is scaled with a blackbody at 5\,000~K.}}
    \label{FigModels}
\end{figure*}

The model from \citet{Compiegne11} (hereafter MC11) is a mixture of PAHs, both neutral and ionized (cations), small and large amorphous carbonaceous grains \citep[SamC and LamC, respectively;][]{Zubko96} with different size distributions, and amorphous silicate grains \citep[aSil;][]{DraineLee84}, i.e. a total of five independent components. In our fitting, we choose to only use a single PAH population, by summing the ionized and neutral species together. Given the shape of the emission spectra from the charged and neutral PAHs, our broad-band observations can not constrain them independently.  We also tie (by summing) the big grains (BGs) together, originally described by both large carbonaceous and amorphous silicates.  At $\lambda \geqslant 250~\mathrm{\mu m}$, the emissivity law of both carbon and silicate grains in this model is the same ($\beta \sim 1.7-1.8$). Hence, they cannot be discriminated from their emission alone and allowing them to vary would result in the fitting arbitrarily choosing one or the other type of grains. Their variations with the temperature are not different enough to be helpful in breaking the degeneracy. In fine, we use three independent grain populations for this model.

The second model we used is the one for the diffuse-ISM type dust in The Heterogeneous Evolution dust Model at the IAS \citep{Jones13, Kohler14}, hereafter THEMIS. In this model, the dust is described by three components, split into four populations: very small grains (VSG) made of aromatic-rich amorphous carbon, large(r) carbonaceous grains with an aliphatic-rich core and an aromatic-rich mantle, and amorphous silicate grains with nano-inclusion of Fe/FeS and aromatic-rich amorphous carbon mantle. The silicate grains are split in two populations: Pyroxene ($\mathrm{-(SiO_3)_2}$) and Olivine ($\mathrm{-(SiO_4)}$). We choose to tie these two silicate population for the same reason than previously mentioned: up to 500~$\mathrm{\mu m}$, they cannot be discriminated only by their emission. We therefore use three independent grain populations for THEMIS.

There is no clear correspondence between the two models because of their different (yet sometimes overlapping) grain type definitions. The PAHs will only be a feature of the MC11 \citep{Compiegne11} model, the SamC will refer to the small-amorphous carbon grains, and BGs will point to the large-amorphous carbon grains and amorphous silicates. In THEMIS \citep{Jones13, Kohler14}, sCM20 and lCM20 will refer to the small- and large- amorphous carbon grains, respectively, and we will refer to the pyroxene (aPyM5) and olivine (aOlM5) grains altogether as aSilM5. Figure \ref{FigModels} shows the models as they were used, i.e. with the tied populations.

The free parameters we allow to vary in our fitting are thus $Y_\mathrm{PAHs}$, $Y_\mathrm{SamC}$ and $Y_\mathrm{BGs}$ in the MC11 model, and $Y_\mathrm{sCM20}$, $Y_\mathrm{lCM20}$, $Y_\mathrm{aSilM5}$ in THEMIS. The $Y_\mathrm{i}$ are scaling factors of the solar neighborhood abundances $M_\mathrm{i} / M_\mathrm{H}$, where $i$ is one of the grain species \citep[e.g.][]{Compiegne11}. The SEDs are scaled through these parameters.
Additionally, the ISRF environment will change with different approaches. This is explained in Section \ref{SecModelComp}.
Finally, due to short wavelengths and a non-negligible emission from stars in the IRAC bands, we also add a stellar component modeled as a black-body spectrum at 5\,000 K. This parameter is scaled through a stellar density $\Omega^*$.

\subsection{DustBFF}
The fitting technique follows the work of \citet{Gordon14} and this ensemble of methods is named DustBFF for Dust Brute Force Fitter. \citet{Gordon14} make use of a multi-variate distribution to determine the probability of a model to fit the data. In this distribution, the $\chi^2$ value is computed from the difference between the model prediction and the data, on which we apply uncertainties through a covariance matrix (equation 18 in their paper). The correlation matrix results from the combination of the uncertainties from the background estimation ($\mathbb{C}_\mathrm{bkg}$) and the errors on the observed fluxes from the instrumentation ($\mathbb{C}_\mathrm{cal} = \mathbb{C}_\mathrm{corr} + \mathbb{C}_\mathrm{uncorr}$).

The background covariance matrix $\mathbb{C}_\mathrm{bkg}$ is calculated from the regions outside of the galaxy mentioned in Section \ref{SecData} (see Eq. 23 in \citet{Gordon14}).

The calibration matrix $\mathbb{C}_\mathrm{cal}$ is determined from the detailed calibration work done for each instrument. For `uncorrelated' errors, we usually refer to the characteristic of repeatability of measurements. This term describes how stable a measurement is in instrument units at high signal-to-noise. This error is not correlated between the different bands of a same instrument, and depicts the diagonal elements of the $\mathbb{C}_\mathrm{uncorr}$ matrix.
The measured gain of an instrument is a `correlated' error. For example, estimating the sky level outside of the bright star (used for calibration) measured relies on various possible methods (e.g. increasing apertures). The systematic errors made in any of the methods propagate throughout the instrument, introducing correlated uncertainties.
We account for calibration uncertainties as correlated errors.
The IRAC and MIPS instruments were calibrated with stars. The IRAC uncertainties were taken from \citet{Reach05}. The instrument has a stability accounting for uncorrelated error of 1.5\%; the absolute calibration leads to uncertainties of 1.8\%, 1.9\%, 2.0\%, 2.1\% at 3.6, 4.5, 5.8, and 8.0~$\mathrm{\mu m}$, respectively.  The MIPS uncertainties were taken from \citet{Engelbracht07MIPS24} and \citet{Gordon07MIPS70}. The repeatability at 24 and 70~$\mathrm{\mu m}$ are 0.4\% and 4.5\%, respectively. Absolute calibrations were made from star observations and give 2\% and 5\% error, at 24 and 70~$\mathrm{\mu m}$, respectively. 
The PACS calibration was done with stars and asteroids models, with an absolute uncertainty of 5\%, correlated between PACS bands, and a repeatability of 2\% \citep{Muller11_PACS, Balog13}.
The SPIRE calibration used models of Neptune with an absolute uncertainty of 4\% and 1.5\% repeatability uncorrelated between bands \citep{Bendo13, Griffin13}. Both of the absolute uncertainties quoted above (5\% and 4\%) were made upon point source calibration. We choose to double all the uncorrelated uncertainties to account for the error on the beam area that arises for extended sources (see the matrices \ref{Uncorr_IRAC} - \ref{Uncorr_SPIRE}.) 

\begin{equation}
    \mathbb{C}{^{\mathrm{IRAC}}_\mathrm{uncorr}} =
    \left(
    \begin{array}{ccccccccccc}
        0.015^2 & 0         & 0         & 0         \\
        0       & 0.015^2   & 0         & 0         \\
        0       & 0         & 0.015^2   & 0         \\
        0       & 0         & 0         & 0.015^2    \\
    \end{array} \right),
\end{equation}
\begin{equation}
    \mathbb{C}{^{\mathrm{MIPS}}_\mathrm{uncorr}} =
    \left(
    \begin{array}{ccccccccccc}
        0.004^2  & 0         \\
        0               & 0.045^2  \\
    \end{array} \right),
\end{equation}
\begin{equation}
    \mathbb{C}{^{\mathrm{PACS}}_\mathrm{uncorr}} =
    \left(
    \begin{array}{ccccccccccc}
        0.02^2  & 0         \\
        0       & 0.02^2    \\
    \end{array} \right),
\end{equation}
\begin{equation}
    \mathbb{C}{^{\mathrm{SPIRE}}_\mathrm{uncorr}} =
    \left(
    \begin{array}{ccccccccccc}
        0.015^2 & 0         & 0         \\
        0       & 0.015^2     & 0         \\
        0       & 0         & 0.015^2   \\
    \end{array} \right) \qquad \mathrm{and}
\end{equation}
\begin{equation}
    \mathbb{C}_\mathrm{uncorr} =
    \left(
    \begin{array}{ccccccccccc}
        \mathbb{C}{^{\mathrm{IRAC}}_\mathrm{uncorr}}    &   &   & (0)         \\
            & \mathbb{C}{^{\mathrm{MIPS}}_\mathrm{uncorr}}  &   &          \\
            &   & \mathbb{C}{^{\mathrm{PACS}}_\mathrm{uncorr}}  &           \\
        (0)    &   &   & \mathbb{C}{^{\mathrm{SPIRE}}_\mathrm{uncorr}}          \\
    \end{array} \right).
\end{equation}

\begin{equation}
    \mathbb{C}{^{\mathrm{IRAC}}_\mathrm{corr}} =
    \left(
    \begin{array}{ccccccccccc}
0.036^2   & 0.036^2     & 0.036^2     & 0.036^2     \\
0.036^2   & 0.038^2     & 0.036^2     & 0.036^2     \\
0.036^2   & 0.036^2     & 0.040^2     & 0.036^2     \\
0.036^2   & 0.036^2     & 0.036^2     & 0.042^2     \\
    \end{array} \right),
    \label{Uncorr_IRAC}
\end{equation}
\begin{equation}
    \mathbb{C}{^{\mathrm{MIPS}}_\mathrm{corr}} =
    \left(
    \begin{array}{ccccccccccc}
        0.04^2  & 0.04^2    \\
        0.04^2  & 0.1^2    \\
    \end{array} \right),
    \label{Uncorr_MIPS}
\end{equation}
\begin{equation}
    \mathbb{C}{^{\mathrm{PACS}}_\mathrm{corr}} =
    \left(
    \begin{array}{ccccccccccc}
        0.1^2   & 0.1^2     \\
        0.1^2   & 0.1^2     \\
    \end{array} \right),
    \label{Uncorr_PACS}
\end{equation}
\begin{equation}
    \mathbb{C}{^{\mathrm{SPIRE}}_\mathrm{corr}} =
    \left(
    \begin{array}{ccccccccccc}
        0.08^2  & 0.08^2    & 0.08^2    \\
        0.08^2  & 0.08^2    & 0.08^2    \\
        0.08^2  & 0.08^2    & 0.08^2    \\
    \end{array} \right) \qquad \mathrm{and}
    \label{Uncorr_SPIRE}
\end{equation}
\begin{equation}
    \mathbb{C}_\mathrm{corr} =
    \left(
    \begin{array}{ccccccccccc}
        \mathbb{C}{^{\mathrm{IRAC}}_\mathrm{corr}}    &   &   & (0)         \\
            & \mathbb{C}{^{\mathrm{MIPS}}_\mathrm{corr}}  &   &          \\
            &   & \mathbb{C}{^{\mathrm{PACS}}_\mathrm{corr}}  &           \\
        (0)    &   &   & \mathbb{C}{^{\mathrm{SPIRE}}_\mathrm{corr}}          \\
    \end{array} \right).
\end{equation}

\subsection{Model (re-)calibration}
\label{SecModelRecalib}
Since we want to investigate the differences of two dust grain models, independently of the fitting algorithm, we find critical to make sure that they share the same calibration. Moreover, this calibration should be made using the same measurements with the same technique. Usually, dust grain models are calibrated to reproduce the diffuse MW dust emission \citep[e.g.][]{Boulanger96} and extinction, with constraints on elemental abundances from depletion measurements \citep[e.g.][]{Jenkins09, Tchernyshyov15}. However, they often do not share the same calibration technique or the same constraint measurements.

MC11 and THEMIS size distributions are calibrated on the diffuse extinction in the MW. Measurements at high Galactic latitude from the Cosmic Background Explorer \citep[COBE;][]{Bennett96}, coupled with the Wilkinson Microwave Anisotropy Probe \citep[WMAP;][]{Jarosik11} and Infrared Space Observatory \citep[ISO;][]{Mattila96} trace the global SED of dust emission. It was correlated with \ion{H}{I} measurements and is expressed in flux units per hydrogen atoms. Hence, in both models, the dust grain `masses' are given as dust-to-hydrogen ratios $M_{\mathrm{dust}}/M_\mathrm{H}$. The total dust mass in each model implies a hydrogen-to-dust ratio that varies from one model to the other. Although each model fits the MW dust emission at high latitude, given their different dust description, they do not necessarily share the same gas-to-dust ratio. However, we do think that this should be a reference point in calibrating dust model, as this can be measured with other methods. In the MW, we follow the value of \citet{Gordon14} and use the diffuse MW hydrogen-to-dust ratio to be 150 (derived from \citet{Jenkins09}, for $F_* \sim 0.36$). 


To ensure that both models produce the same result when fit to the MW diffuse ISM, we ``recalibrate'' the models using the ISO, COBE and WMAP measurements of the local ISM as described in \citet{Compiegne11}. We do not take into account the 0.77 correction for the ionized gas, in order to be consistent with the depletion work of \citet{Jenkins09} which does not correct the ionized gas contribution.  We integrated this spectrum in the \textit{Spitzer} and \textit{Herschel} photometric bands and obtain an SED whose values are shown in Table \ref{TableDiffISM}. The PACS and SPIRE values are very close to the ones displayed in \citet{Gordon14} (Section 5.1 in their paper).

We use the DustBFF fitting technique to scale the full spectrum of each model to the SED described in Table \ref{TableDiffISM} and find the factor that gives the adopted gas-to-dust ratio of 150. We do not allow the grain species to vary from one another, and we choose to keep the same ratios between population as described by the model. We set the ISRF at $U=1$, i.e. the same used for the model definition. The fits thus consist in adjusting the global emissivity, and scale the \emph{total} emission spectrum. We build a different correlation matrix for the estimated flux uncertainties from the observing instruments quoted. Following \citet{Gordon14} we assumed 5\% correlated and 2.5\% uncorrelated uncertainties at long wavelengths for the COBE, FIRAS and DIRBE instruments (accounting for PACS and SPIRE bands). We presume a 10\% error for both correlated and uncorrelated uncertainties at short wavelengths given the resolution of ISO (accounting for IRAC and MIPS bands). We derive a scaling factor which is the result of the fit of the models. The final correction factors are 1.6 and 2.42 for the whole spectrum of THEMIS and the MC11 model, respectively. 
These factors aim at self-calibrating the models to give the same gas-to-dust ratio of 150 for the same MW SED. We find this step crucial as our goal is to compare two models, independently of the fitting method, for which differences are eliminated by the use of a common fitting procedure. It should be noted that, if the first step aims at a rigorous fit to the MW SED, the second step's goal is to adjust the GDR and therefore moves away from a good fit.

We convert the emission output from DustEM $4\pi \, \nu I_\nu$ (in $\mathrm{erg~s^{-1}~cm^{-2}~(H~cm^{-2})^{-1}}$) to surface brightness ($\mathrm{MJy~sr^{-1}}$), with the scaling factors as following:
\begin{equation}
    S_\lambda = 4\pi \ \nu I_\nu \times 2.65 \ 10^{21} \times \lambda \times 
    \begin{cases}
    1.6 \; \textrm{if we use THEMIS} \\
    2.42 \; \textrm{if we use MC11}
    \end{cases}
\end{equation}

\renewcommand{\arraystretch}{1.1}
\begin{table}[]
    \caption{The values of the local diffuse ISM integrated in the \textit{Spitzer} and \textit{Herschel} bands, used for calibration, in MJy~$\mathrm{sr^{-1}}$ $\times 10^{20}$~H~atom$^{-1}$.}
    \centering
        \begin{tabular}{cc|cc}
        \hline
        \hline
        Bands & Diffuse ISM & Bands & Diffuse ISM \\
        \hline
        IRAC3.6 & 0.00235 & PACS100 & 0.714 \\
        IRAC4.5 & 0.00206 &  PACS160 & 1.55 \\
        IRAC5.8 & 0.0134 &  SPIRE250 & 1.08 \\
        IRAC8.0 & 0.0431 &  SPIRE350 & 0.561 \\
        MIPS24 & 0.0348 &   SPIRE500 & 0.239 \\
        MIPS70 & 0.286 & & \\
        \hline
        \hline
        \end{tabular}
    \label{TableDiffISM}
\end{table}

\section{Model comparison}
\label{SecModelComp}
We vary a number of dust model parameters that affect the SED shape and fit these new spectra to the data. In this study, we mainly examine dust emission when it is illuminated by different ISRF mixtures. At higher ISRFs, we expect the IR peak to shift to shorter wavelengths. We focus on this behaviour after considering the shape of global SEDs in the MCs.
In all cases, we vary the $Y_{\mathrm{i}}$ parameters, which adjust each grain abundance. We also change the ISRF intensity, scaled by the free parameter $U$. In the whole study, we use the standard radiation field defined by \citet{Mathis83}.  The $U=1$ case corresponds to the solar neighborhood ISRF $\mathrm{U_\odot}$.
In one case, we vary the small grain size distribution. Throughout the fitting, we do not change the large grain size distributions, and therefore assume no change between the MW and the MCs, regarding this aspect.

We choose to fit each pixel that is detected at least $3\sigma$ above the background 8 bands (IRAC8.0, MIPS24 and MIPS70, PACS100 and PACS160 and all SPIRE bands): we do not impose the detection condition at 3.6, 4.5 and 5.8~$\mathrm{\mu m}$ as these bands can include a significant contribution from stars. We do include all the IRAC observational data in the fitting. In the following, the `faint' or `bright' aspect of a pixel is based on its emission at 500~$\mathrm{\mu m}$.

The main output of DustBFF is the $n$D likelihood function \citep[][]{Gordon14}. From this, we can approach the ``best-fit'' value with different estimators. First, we use the `max' value, defined as the maximum likelihood, or sometimes called traditional $\chi^2$. It reflects the closest model to the observations, and we use it for residual calculations. The residuals illustrate how the models match the data and are expressed as the error $(\mathrm{data}-model)/model$. Another way to derive results from the likelihood function is to randomly sample it, reflecting its shape and the fitting noise. We use this ``realization'' method to derive dust masses.

\begin{figure*}
   \centering
       \includegraphics[width=0.5\textwidth]{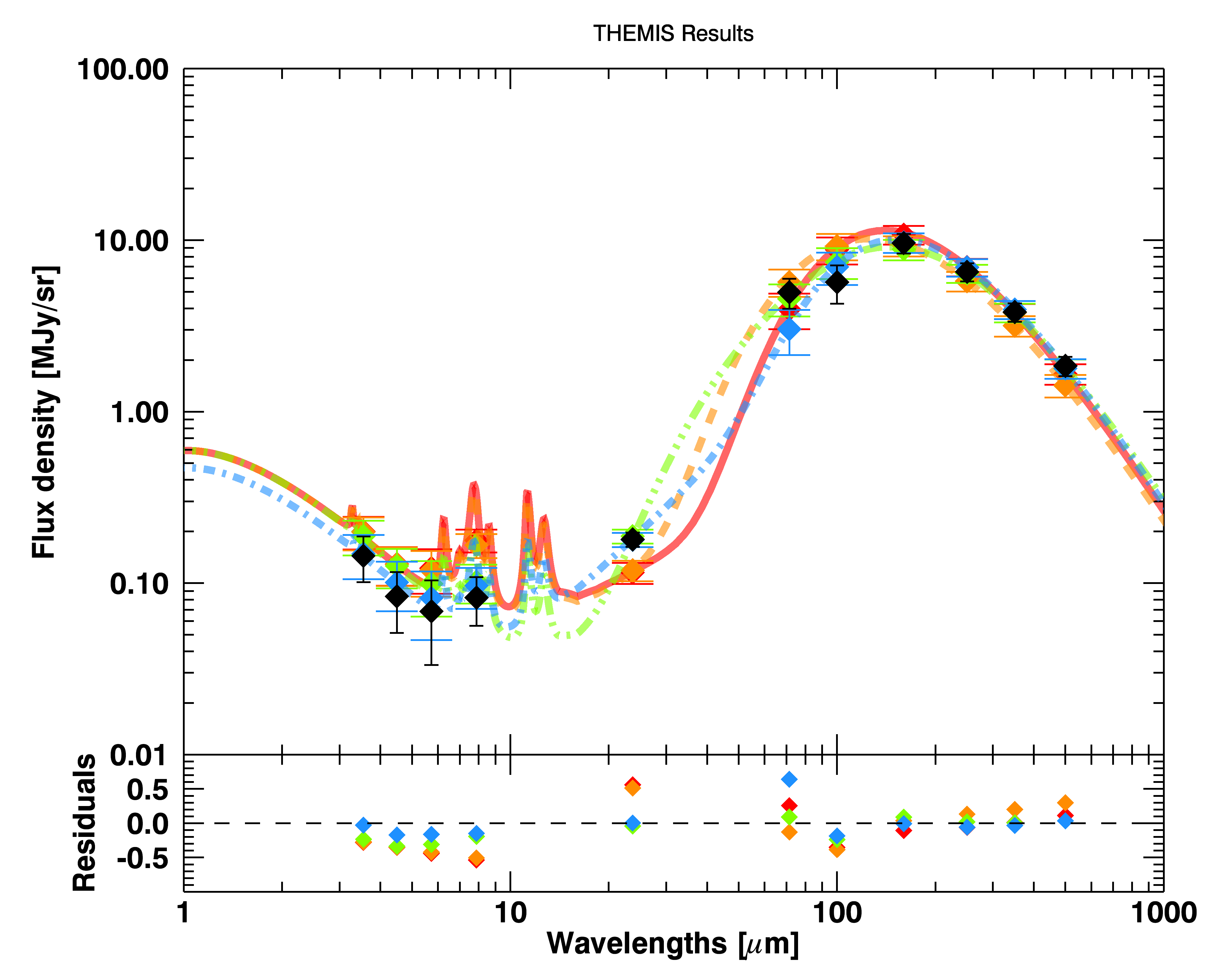}\hfill\includegraphics[width=0.5\textwidth]{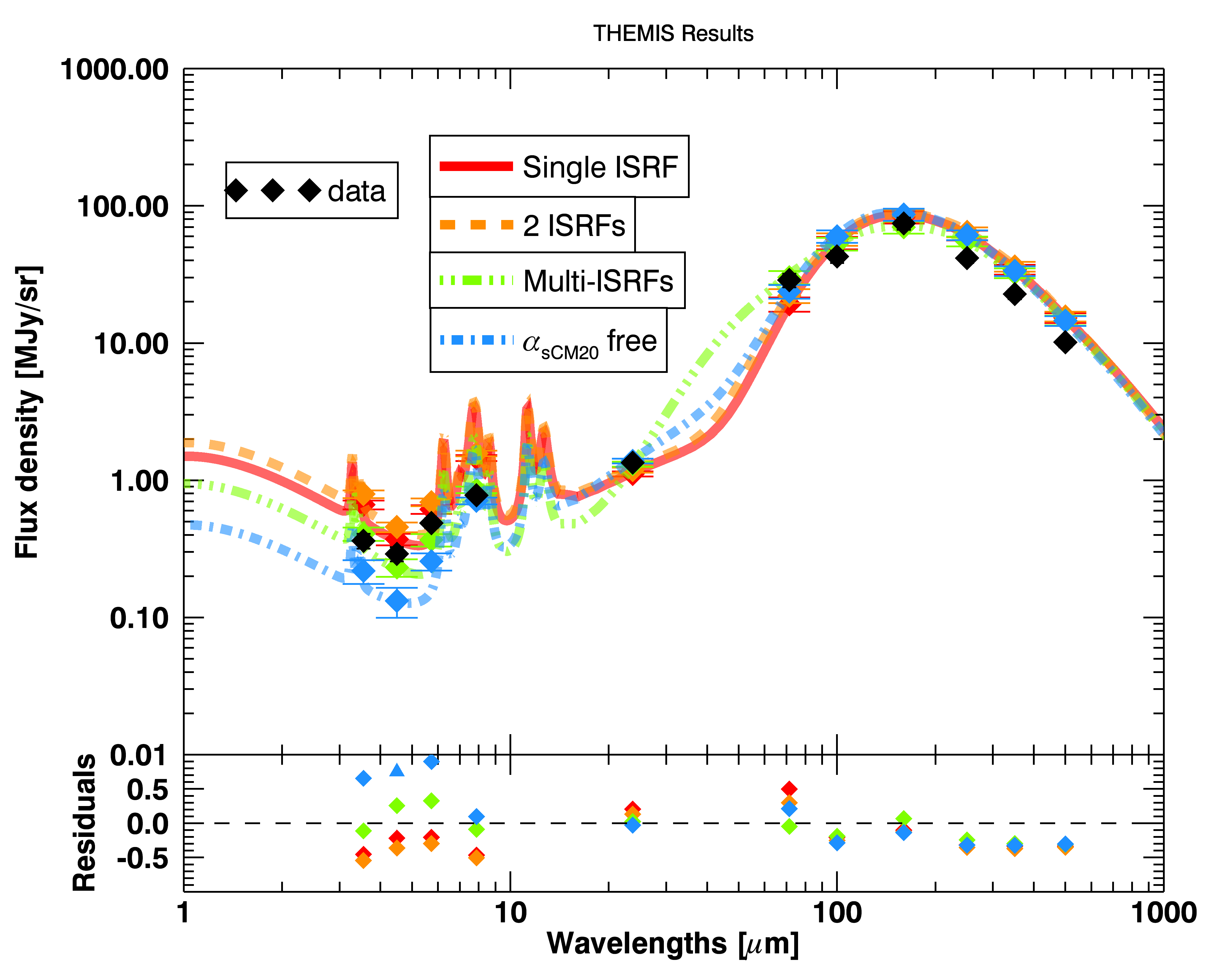}
       \caption{\small{Examples of the fits results in the SMC for a faint (left) and a bright pixels (right). We note the difference in the residuals, namely in the FIR, best fit in faint (diffuse) environments. We also see the impact of a change in the 8/24~$\mathrm{\mu m}$ slope on the fits in the MIR.}}
       \label{FigSMCFits}
\end{figure*}

Figure \ref{FigSMCFits} shows fitting results for two pixels in the SMC, one with faint emission (left) and one with bright emission (right), for THEMIS only. This figure gives an idea of the different model variations that we describe in the following sections.

\subsection{Single ISRF}
\label{SecSingleISRF}
We first used the models with a single ISRF environment. This simply means that each spectrum is calculated from the emission of grains illuminated by a single ISRF, the strength of which varies. We do not change the shape or hardness of the ISRF. 
In Figure \ref{FigLMC_AJMCs}, we show the distribution of fractional residuals expressed as $(\mathrm{data}-model)/model$, in the SMC (top) and the LMC (bottom) for the two different models. The red line shows the results for THEMIS and the purple-dot line, the MC11 model.

First, the residuals do not have a Gaussian shape. In some bands (e.g. PACS160 in the upper image of Figure \ref{FigLMC_AJMCs}), the residuals have a large negative tail.

Second, the large grain population (aSil+LamC tied) in the MC11 model does not reproduce well the FIR emission at $\lambda \geqslant 100~\mathrm{\mu m}$ in the SMC, where the fractional residual distribution is broad. THEMIS with a single ISRF, on the other hand, seems to reproduce the long-wavelength part of the SED in the SMC better than the MC11 model. We can notice that the model is still, on average, slightly too high to properly reproduce the observations (noticeable by a mean of the residuals below 0), in the SMC and the LMC.

The FIR slope of the big grains in the MC11 model, described by a $\beta \sim 1.7 - 1.8$, is not compatible with the observed SEDs in the SMC, which show peculiarities: flat FIR emission, broad IR peak. The modeled slope is too steep to reproduce the flatter emission spectrum observed below $500~\mathrm{\mu m}$. Another explanation of the broad residuals may come from the ratio between silicate and carbonaceous material. This ratio is believed to be fairly constant across the Galaxy. Tying {aSil+LamC} as one component implies that this ratio imposed by the original model is kept throughout the fitting. 
Further tests showed that the initial assumption of tying these two populations is justified and does not prevent a better fit to the long-wavelength observations. In THEMIS, the large carbonaceous grain emission in particular exhibits a flatter FIR slope than the MC11 model (see Figure 1). This is likely why THEMIS reproduces the FIR SED better and is likely the reason of a better reconstruction of the observations. 

The excesses visible at 70 and $100~\mu$m with the MC11 model are better fit with the THEMIS model.
At short wavelengths, and especially at $8~\mathrm{\mu m}$, the MC11 model shows smaller residuals than THEMIS.
This is likely the consequence of an additional degree of freedom in that part of the spectrum. THEMIS uses a single population to depict the small grains emission, whereas MC11 uses two distinct grain species (PAHs and SamC, Figure \ref{FigModels}).

A single ISRF is arguably not a good reproduction of the physical environment of dust and the nature of the observations. Mixture of the starlight along the line of sight is likely to occur. A single ISRF remains nonetheless the simplest model and can be used to compare to more simple models like SMBB or BEMBB, as they only assume a single ISRF heating as well. Figure 4 of \citet{Gordon14} shows the residuals at $250~\mathrm{\mu m}$. On average, the BEMBB model (the one they retain as best in their study) gives better residuals than our fitting. In both cases, we can notice a slight shift toward negative values, indicating the BEMBB model is too high with respect to the observations. Yet, their results better match the data. This is likely due to the fact that the FIR slope can be directly adjusted using the $\beta_2$ parameter, independently in each pixel.

\subsection{Mixtures of ISRFs}
\label{SecMultiISRF}
The next level of complexity for the heating environment is to use two ISRFs. In this case, we consider two components of dust: we calculate the emission of each grain population when irradiated by two ISRFs with different strengths, which leads to a ``warm dust'' and a ``colder one'', and then mix the spectra with a fraction $f^{\mathrm{warm}}$: 
\begin{equation}
    I_\nu = \sum_X Y_{\mathrm{X}} \left ( f^{\mathrm{warm}} 
                        I_\nu^{\mathrm{X^\mathrm{warm}}} + (1-f^{\mathrm{warm}})
                        I_\nu^{\mathrm{X^\mathrm{cold}}} \right ),
\end{equation}
where $\mathrm{X=\{aSilM5; lCM20; sCM20\}}$ (THEMIS). The fraction parameter $Y_{\mathrm{X}}$ is identical for all grain populations. Effectively, we have two parameters $U^\mathrm{warm}$ and $U^\mathrm{cold}$, that both scale up and down the ISRF.
It physically means that we model two dust masses $M^\mathrm{warm}_\mathrm{dust}$ and $M^\mathrm{cold}_\mathrm{dust}$, instead of a single effective dust mass as in Section \ref{SecSingleISRF}.
The $I_\nu^{\mathrm{X^\mathrm{warm}}}$ and $I_\nu^{\mathrm{X^\mathrm{cold}}}$ refer to the dust SEDs heated by $U^\mathrm{warm}$ and $U^\mathrm{cold}$, respectively, with $U^\mathrm{cold} < U^\mathrm{warm}$.
\citet{Meisner15} used a similar approach to fit the Planck HFI all-sky maps combined with IRAS 100~$\mathrm{\mu m}$. They showed that this provides better fits in the wavelength range (100 - 3\,000~$\mathrm{\mu m}$) than a simple modified blackbody.

Finally, one can use a more complicated combination of ISRFs. Thus, we also follow the work of \citet{Dale01} in which the final SED is a power-law combination of SEDs at various ISRFs, integrated over a range of strength:
\begin{equation}
    dM_\mathrm{d}(U) \propto U^{-\alpha_{\mathrm{ISRF}}}dU, \qquad 10^{-1}\mathrm{U_{\odot}} \leq U \leq 10^{3.5}\mathrm{U_{\odot}}
\end{equation}
The $\mathrm{\alpha_{ISRF}}$ coefficient is the parameter that regulates the weight of strong/weak ISRFs in the mixture used to irradiate the dust in a multi-ISRFs model. A low $\mathrm{\alpha_{ISRF}}$ gives more weight to the high ISRFs. We allow the $\mathrm{\alpha_{ISRF}}$ parameter to vary between 1 and 3, as suggested by previous studies \citep[e.g.][]{Bernard08}.

In Figure \ref{FigSMCFits}, we note that the use of multiple ISRFs leads to a better match of the $24~\mu m$ data. In the faint pixel, the emission in the IRAC bands is dominated by starlight and not extremely sensitive to small carbon grains, except at $8.0~\mu m$. The differences in the fits in a faint or bright pixel could mean that diffuse regions are better reproduced by THEMIS than brighter regions, which are most likely denser. In these regions, dust may be significantly different in terms of dust properties, and a fixed dust grain model may not be appropriate.

\begin{figure*}
    \centering
        \includegraphics[width=0.6\textwidth]{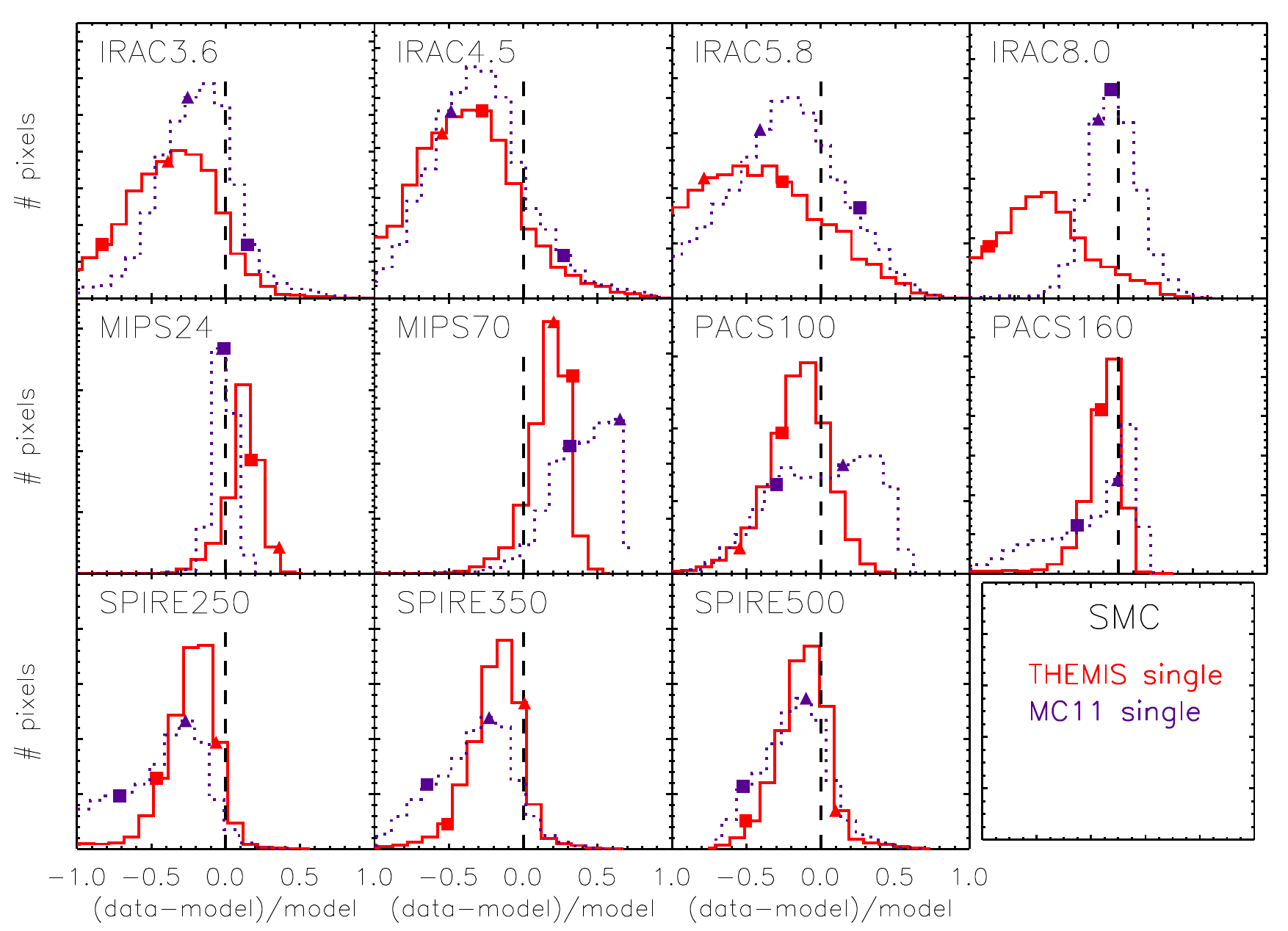}
        \includegraphics[width=0.6\textwidth]{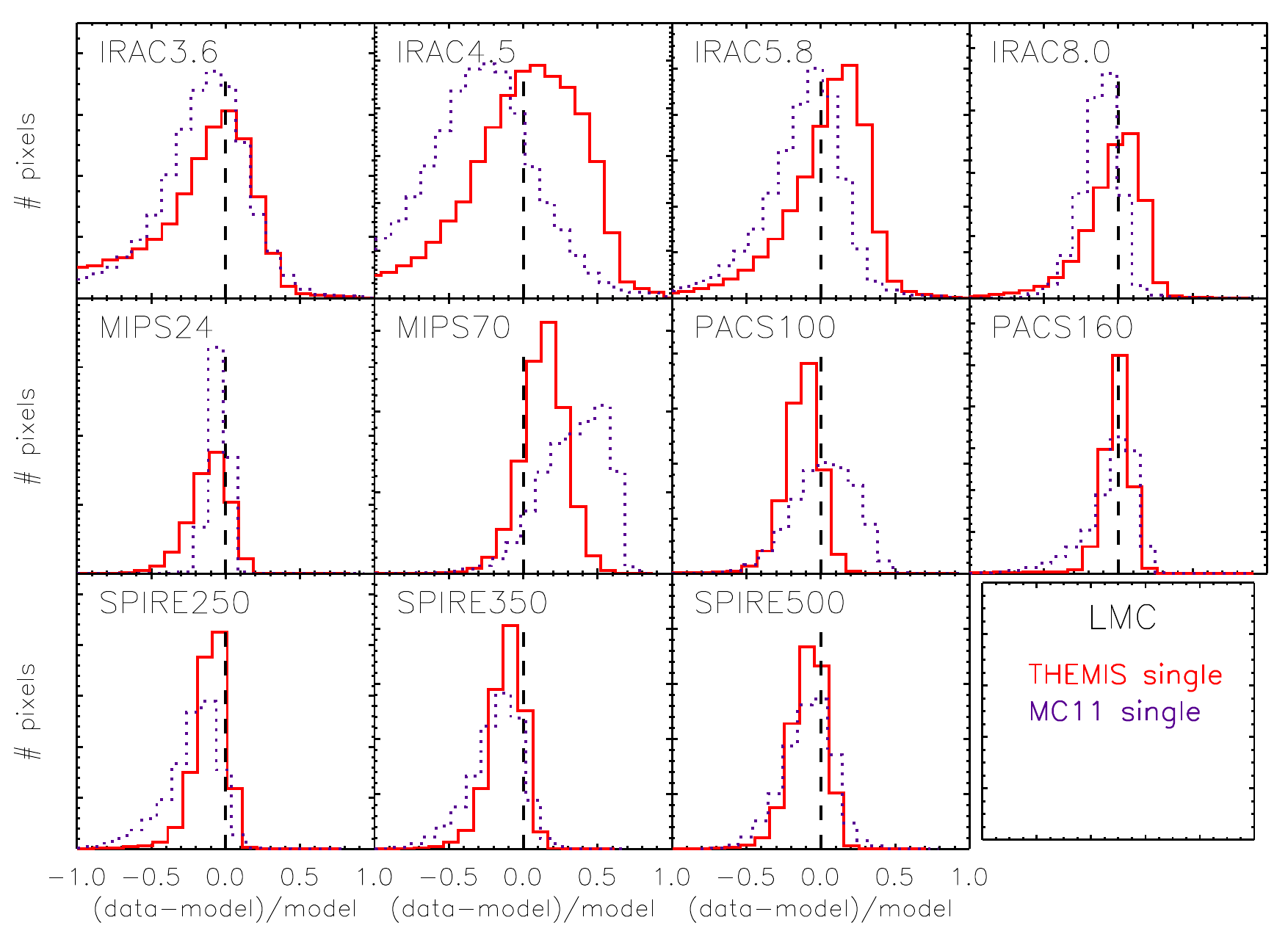}
    \caption{\small{Histograms of fractional residuals for the MC11 (purple-dot) and THEMIS (red) models in a single ISRF environment, in the SMC (top), and the LMC (bottom). On the upper panel, triangles and squares show the residuals for the same faint and bright, respectively, pixels than Figure \ref{FigSMCFits}; colors correspond to the models.}}
    \label{FigLMC_AJMCs}
\end{figure*}

\begin{figure*}
    \centering
        \includegraphics[width=0.7\textwidth]{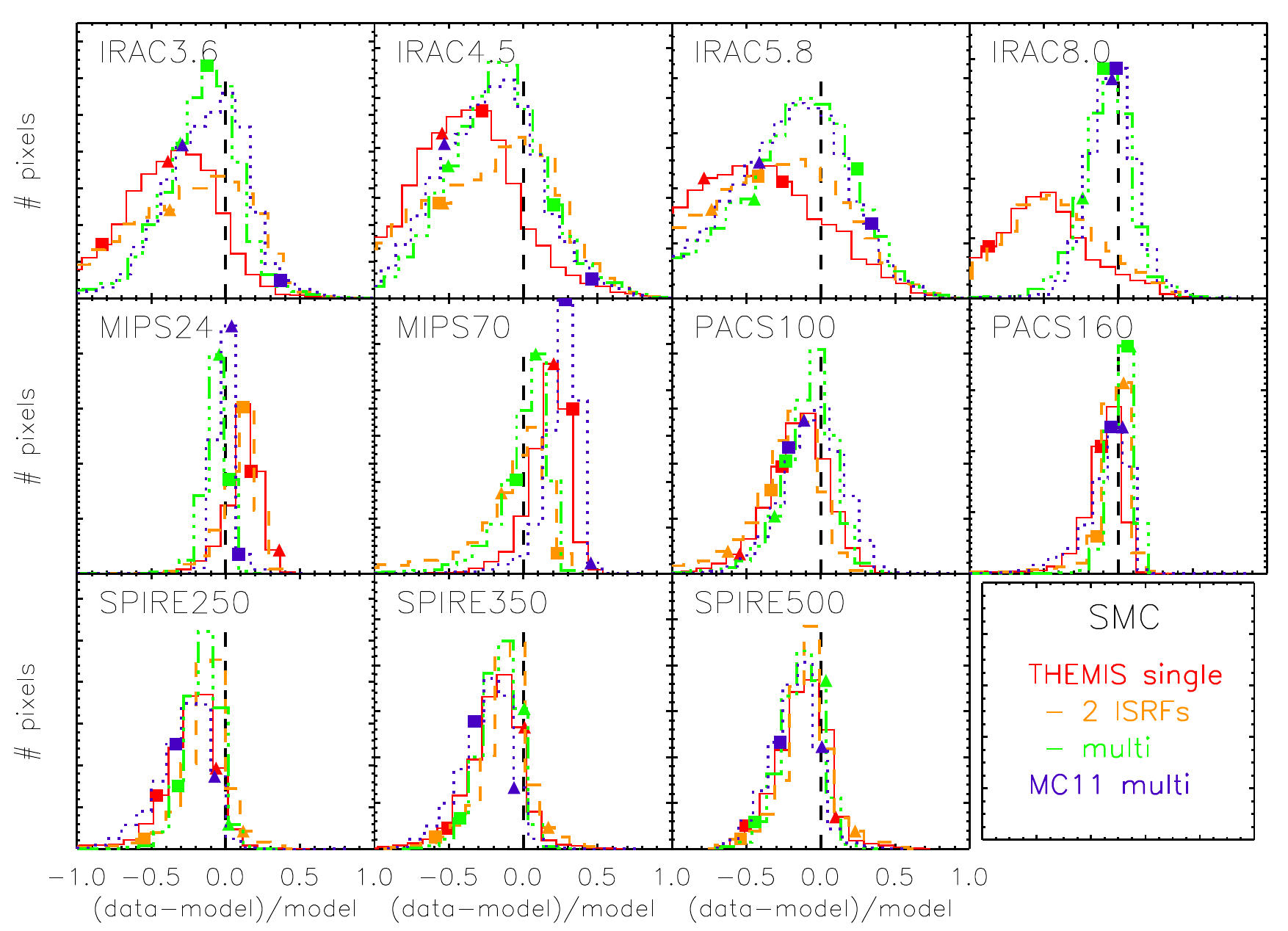}
        \includegraphics[width=0.7\textwidth]{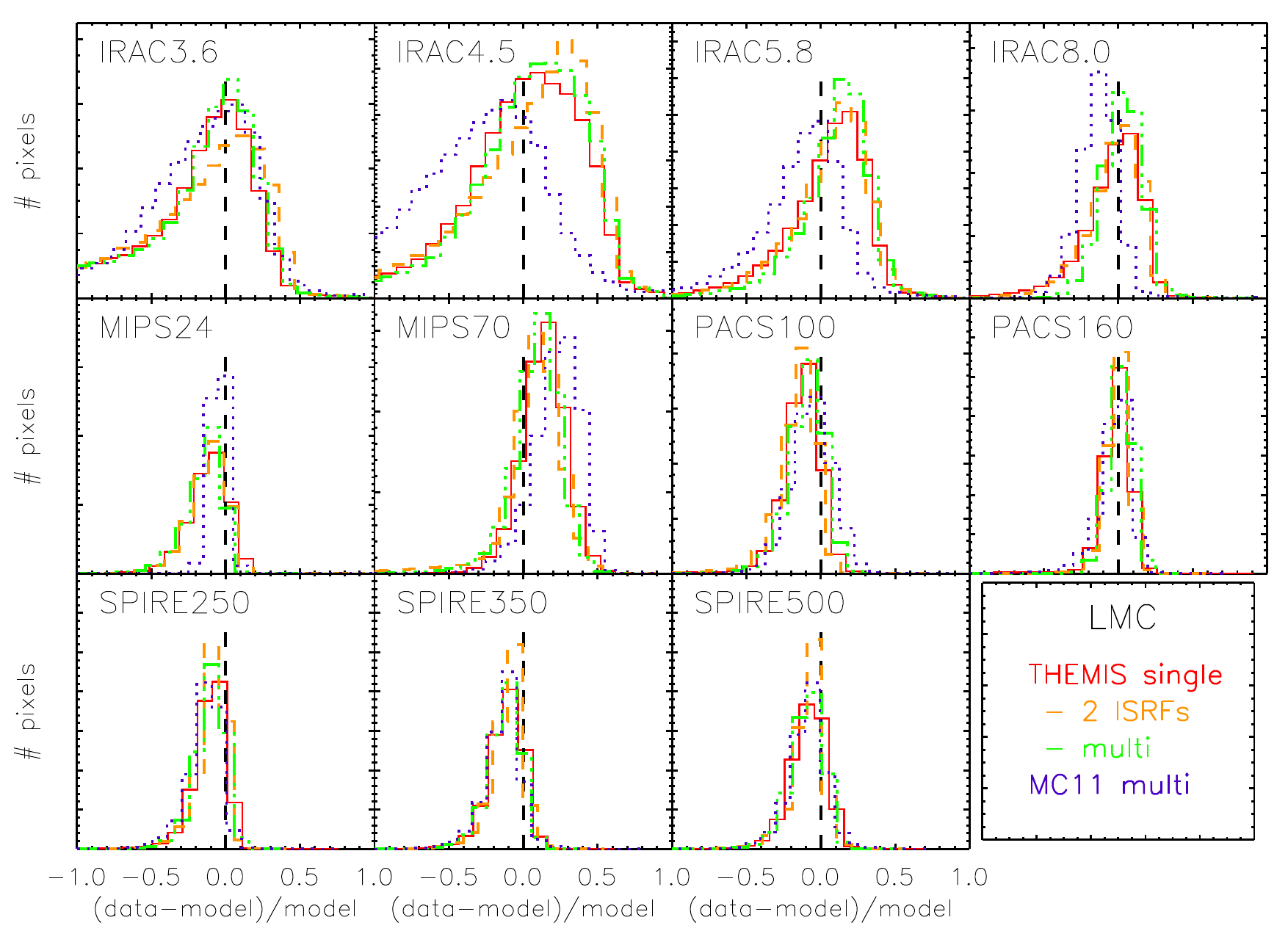}
    \caption{\small{Histograms of fractional residuals for THEMIS in a 2-ISRFs (orange-dashed) and a multi-ISRFs (green-dash-triple dot) environments, in the SMC (top), and the LMC (bottom). For reference, THEMIS in a single ISRF is shown in red line. The MC11 model in a multi-ISRFs model is pictured, but does not exhibit strong improvement compared to a single ISRF environment.}}                            
    \label{FigLMC_AJstm}
\end{figure*}

In Figure \ref{FigLMC_AJstm}, we show the residuals for THEMIS used in a 2 ISRFs environment (orange-dashed line), and THEMIS and MC11 model in a multi-ISRFs environment (green-dash-triple dot line). As a reference, we keep the results for the simplest THEMIS model (i.e. ``single ISRF''; red line).
The FIR residuals for the MC11 model do not show improvements with respect to those of a single ISRF environment (Figure \ref{FigLMC_AJMCs}, purple-dot line). At $\lambda \leqslant 24~\mu m$, it follows THEMIS with the same environment, and hence does not have strong assets.
At long wavelengths ($\lambda \geqslant 100~\mathrm{\mu m}$), THEMIS in the two environments described in this section has residuals centered on 0, and are no longer shifted below 0 as the single ISRF model. This is most visible in the SMC (top panel). In the LMC (bottom panel), the single ISRF model provides a fairly good fit, and the improvements of the other models are less significant.

We can see the multi-ISRF model improves the fits at $5.8~\mathrm{\mu m} \leqslant \lambda \leqslant 70~\mathrm{\mu m}$, in both the SMC and LMC. Mixing the dust heated by different ISRFs notably helps to match the data at 8.0 and 24~$\mu m$. Using only two ISRFs components does not seem enough, and this model reproduces the same SED than a single ISRF model at these wavelengths.
The efficiency of using a power-law is due to its effect on the 8/24~$\mu m$ slope. By steepening it, it matches both the NIR and MIR data better.

After this section, we no longer use the MC11 model. It suffers from strong divergence with the data and the effects brought by using more than a single ISRF do not improve the quality of the fits.

\subsection{Varying Small Grains}
We saw in Section \ref{SecSingleISRF} that a single ISRF does not match very well the data at short wavelengths: the residuals are broad and mostly negative (i.e. the model is too high with respect to the observations). In Section \ref{SecMultiISRF}, we tested different environment changes to try to better account for the shape of SEDs. But variations in grain size distribution can have also an impact on the shape of the dust emission.
In THEMIS, the small grain size distribution is described by a power-law, partly defined as ${dn/da} \propto a^{-\alpha_\mathrm{sCM20}}$, where $a$ is the grain radius. In order to obtain better fits at these wavelengths (3.6, 4.5, 5.8, 8.0, 24~$\mathrm{\mu m}$), we investigate the impact of changing the sCM20 size distribution. In this approach, we keep a single ISRF but allow the $\alpha_\mathrm{sCM20}$ parameter to vary.

\begin{figure*}
    \centering
        \includegraphics[width=0.7\textwidth]{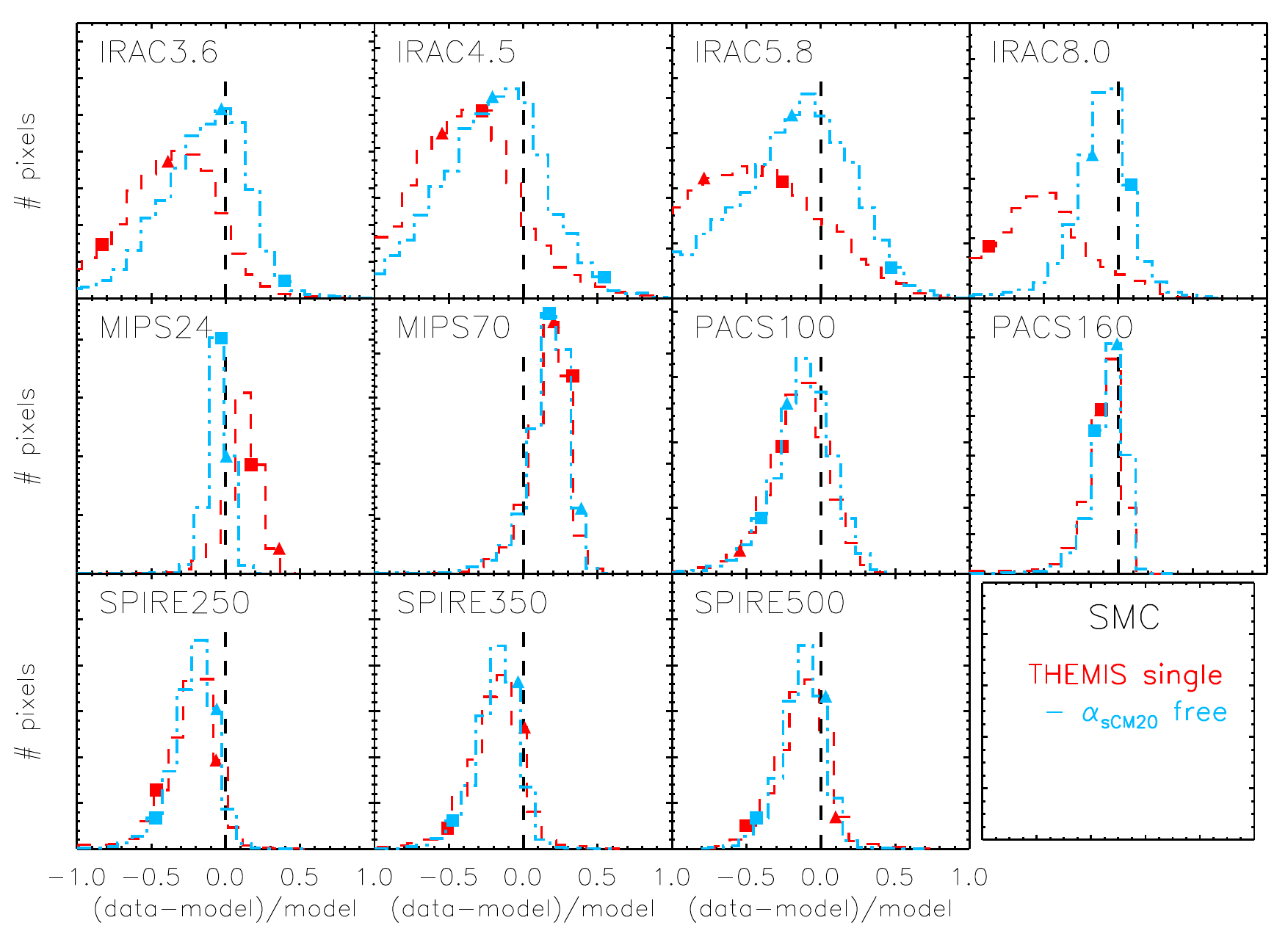}
        \includegraphics[width=0.7\textwidth]{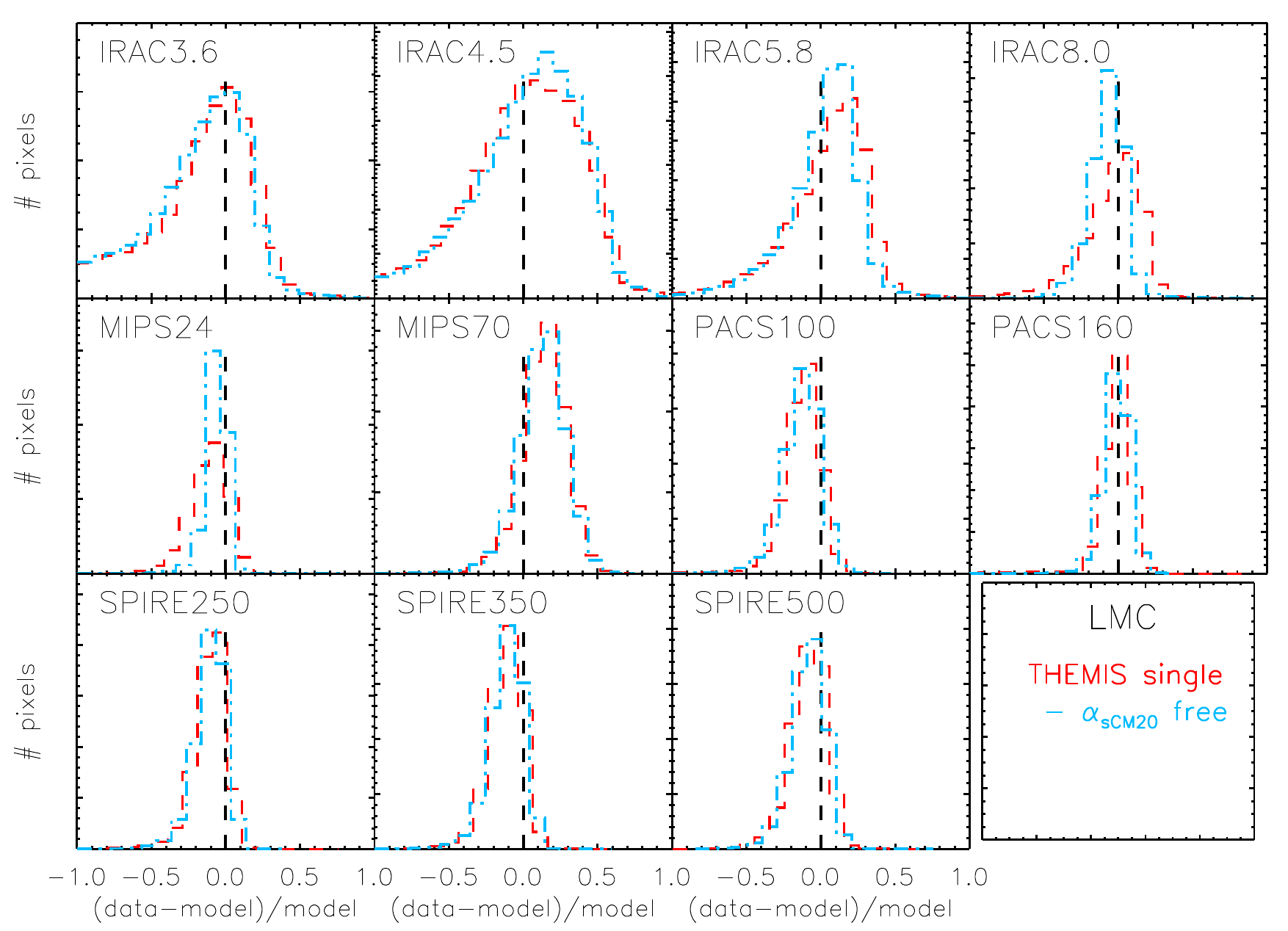}
    \caption{\small{Histograms of fractional residuals for THEMIS in a single ISRF environment with a change of the sCM20 size distribution (light-dash-dot blue), in the SMC (top), and the LMC (bottom). For reference, THEMIS in a single ISRF is shown in red line.}}
    \label{FigLMC_AJsa}
\end{figure*}

In Figure \ref{FigLMC_AJsa}, we show the residuals for this variation (light-dash-dot blue) in the SMC (top), and the LMC (bottom). In both galaxies, the fits at $\lambda \geqslant 70~\mathrm{\mu m}$ are not improved by this model compared to a simple single ISRF (red line). However, the residuals show that this model matches the data better at short wavelengths, particularly in the SMC. The ``means'' of the residuals are centered on 0 and the residuals are less broad.
This improvement is due to the change in the shape of the SED brought by the free parameter $\alpha_\mathrm{sCM20}$. When $\alpha_\mathrm{sCM20}$ decreases, the 8/24~$\mathrm{\mu m}$ slope steepens. This helps decreasing the model values in the MIR. On a more physical aspect, when $\alpha_\mathrm{sCM20}$ is lower, the sCM20 mass distribution is rearranged and it leads to fewer very small grains.

The SMC and the LMC exhibits two different fitting results to the $\alpha_\mathrm{sCM20}$ parameter, which can vary between 2.6 and 5.4. In the LMC, we find $\langle \alpha_\mathrm{sCM20} \rangle \sim 5.0$, i.e. the default value set in the THEMIS model to reproduce MW dust emission. In the SMC, we find $\langle \alpha_\mathrm{sCM20} \rangle \sim 4.0$, with $\alpha_\mathrm{sCM20} < 4.0$ in bright regions (e.g. N66, N76, N83) or \ion{H}{II} regions (e.g. S54). The improvement in the residuals comes from a better fit in these regions allowed by a different SED shape in the IRAC and MIPS24 bands. \citet{Bernard08} found that changing the power-law coefficient of the VSG of the \citet{DBP90} model from 3.0 to 1.0 helps matching the data and decrease the $70~\mu$m excess, and our estimate goes in the same direction.

\section{Dust properties inferred from modeling}
\label{SecResults}
Based on residual study (described in Section \ref{SecModelComp}), we choose to investigate dust properties in the Magellanic Clouds as inferred from only THEMIS: with a 2 ISRFs environment, a multi-ISRFs environment, and with $\alpha_\mathrm{sCM20}$ free, in a single-ISRF environment.
Compared to these models, the use of the simple THEMIS or the MC11 model
does not allow a good quality of fits on the MCs. We decide to not use them to derive dust properties and spatial variations.

\subsection{Spatial variations}
\label{SecSpatialVariations}
We investigate the spatial variations of the parameter distribution by building parameter maps. The maps show strong differences from a model to another. 

\begin{figure*}
    \centering
        \includegraphics[width=0.75\textwidth]{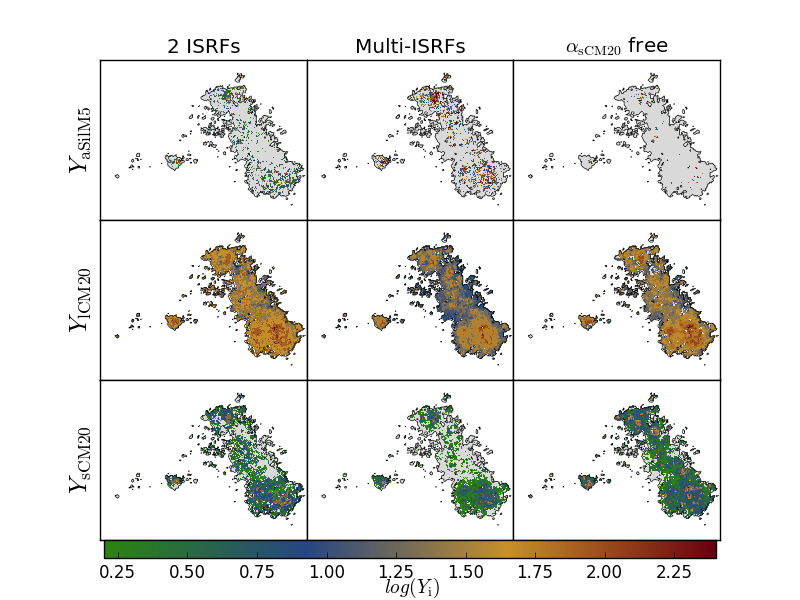}
        \includegraphics[width=0.75\textwidth]{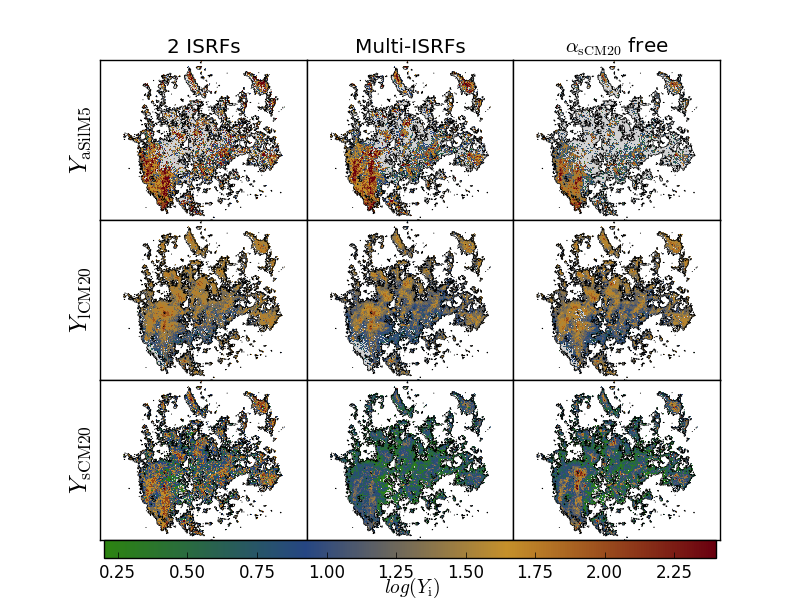}
    \caption{\small{Parameter maps from THEMIS fits, of $Y_\mathrm{aSilM5}$, $Y_\mathrm{lCM20}$, and $Y_\mathrm{sCM20}$ (first, second and third rows), for the 2-ISRFs, multi-ISRFs, and $\alpha_\mathrm{sCM20}$ free (first, second and third column) models. The grey background represent all fit pixels. We notice strong discrepancies from a model to another. The spatial variations are dependent on the dust heating environment, especially for the silicate and small carbonaceous grain components.}}
    \label{FigSMCParamMaps}
\end{figure*}

In Figure \ref{FigSMCParamMaps}, we show the maps of the resulting scaling factors $Y_\mathrm{i}$ for i=\{aSilM5, lCM20 and sCM20\} (first, second and third rows, respectively), in the SMC (top) and the LMC (bottom), for the 2 ISRFs, multi-ISRFs, and $\alpha_\mathrm{sCM20}$ free models (first, second, and third columns, respectively). 
The first and third rows ($Y_\mathrm{aSilM5}$ and $Y_\mathrm{sCM20}$) show the most striking variations. 
We only display meaningful pixels in color, i.e. pixels where the result is higher than its uncertainty. For example, in the upper-right corner of Figure \ref{FigSMCParamMaps} (top), the very few pixels displayed are the only significant pixels. We emphasize that the same pixels were fit in each case, and the discrepancies in the images come from variations in results. As a guide, the faint grey patterns show all fitting pixels.

The silicate fitting, discussed in the next section (\ref{SecSilicates}), is strongly affected by the choice of the heating environment, in both galaxies, although it is particularly noticeable in the SMC. Using a multi-ISRFs model reduces the number of poorly-constrained fits (i.e. with an upper limit). On the other hand, the $\alpha_\mathrm{sCM20}$ free model leaves an extensive portion of the galaxy with unconstrained fits (not-shown pixels, with an upper limit on silicate abundances, i.e. a large uncertainty). The results in the LMC appears to be less variable from a model to another. As seen from the residuals, the models match the LMC observations better than the SMC.
We think this difference comes from the constraints put on the silicate spectrum, which seems to vary from a model to another. In the case of a multi-ISRFs model, the emission of silicates is significant at $70~\mathrm{\mu m}$. The total flux in this band (MIPS70) has thus a stronger silicate contribution. This means the silicate spectrum has one more constraint, at a shorter wavelength. This could be the reason of a better fitting result, compared to other models (where the MIPS70 bands is mostly constrained by smaller grains). The more ``constant'' results in the LMC likely come from the fact that they are closer to MW SEDs, upon which the dust grain models are calibrated.

The $Y_\mathrm{lCM20}$ fitting results (second rows) do not show strong variations from a model to another. In the SMC, all the pixels are fitted, and the discrepancies are ``true'' fitting results. In the LMC, the results are once again less variable and seem to be trustworthy.
We compared our resulting fitting parameters maps to those derived by \citet{Paradis09}. They used the \citet{DBP90} model to fit the \textit{Spitzer} emission of the MCs. It should be noted that their study and ours do not use the same model nor the same fitting technique. They found that the $Y_\mathrm{PAH}/Y_\mathrm{BG}$ ratio is higher in the LMC bar, in both cases with a single ISRF and a multi-ISRFs model. Such a behavior does not appear in our maps. We do not find any spatial trend in the distribution of $Y_\mathrm{sCM20}/(Y_\mathrm{aSilM5}+Y_\mathrm{lCM20})$. However, it is difficult to rigorously establish a comparison as even the grain species are not defined in the same way in the different studies.

The $Y_\mathrm{sCM20}$ fitting results are sensitive to the model. The results from the $\alpha_\mathrm{sCM20}$ free model shows regions with more sCM20 that can be correlated to some point with \ion{H}{II} regions, traced by H$_\alpha$ \citep[][]{Gaustad01}.  
The distribution of the $Y_\mathrm{sCM20}$ parameter in the last column is due to the change of size distribution of the small grain. Changing the power-law coefficient of the sCM20 size distribution has one main advantage: it steepens the 8/24~$\mathrm{\mu m}$ slope. This helps fitting the 8 and $24~\mathrm{\mu m}$ bands in the SMC, as shown in Figure \ref{FigLMC_AJsa}. However, it raises the IR emission peak of the small grains. In regions where $\alpha_\mathrm{sCM20}$ is very low, the IR peak can be fitted by the sCM20 species, and requires only a small contribution of the large grains.

\subsection{Silicate Grains}
\label{SecSilicates}
In Section \ref{SecSpatialVariations}, we saw that the silicate grains component is highly model-dependent, and that most of the pixels are not fit with a reliable uncertainty (Figure \ref{FigSMCParamMaps}). The figure shows that the SMC and the LMC do not exhibit the same results, and that the SMC is more sensitive to the model than the LMC.

In both galaxies, we find pixels that show a likelihood where the silicate component is only constrained as an upper limit (i.e. all models below a given abundance of silicates have the same probability). From the likelihoods in the pixels with an unconstrained value of the silicate component, we can quote a 3-$\sigma$ upper limit for the \emph{absence} of the silicate in the fitting. This upper limit is $Y_\mathrm{{aSilM5}} \sim 10^{0.4} M{^\mathrm{aSilM5}}{_\odot} / M_\mathrm{H}$, in both the SMC and the LMC. 

\begin{figure}
    \centering
        \includegraphics[width=0.45\textwidth]{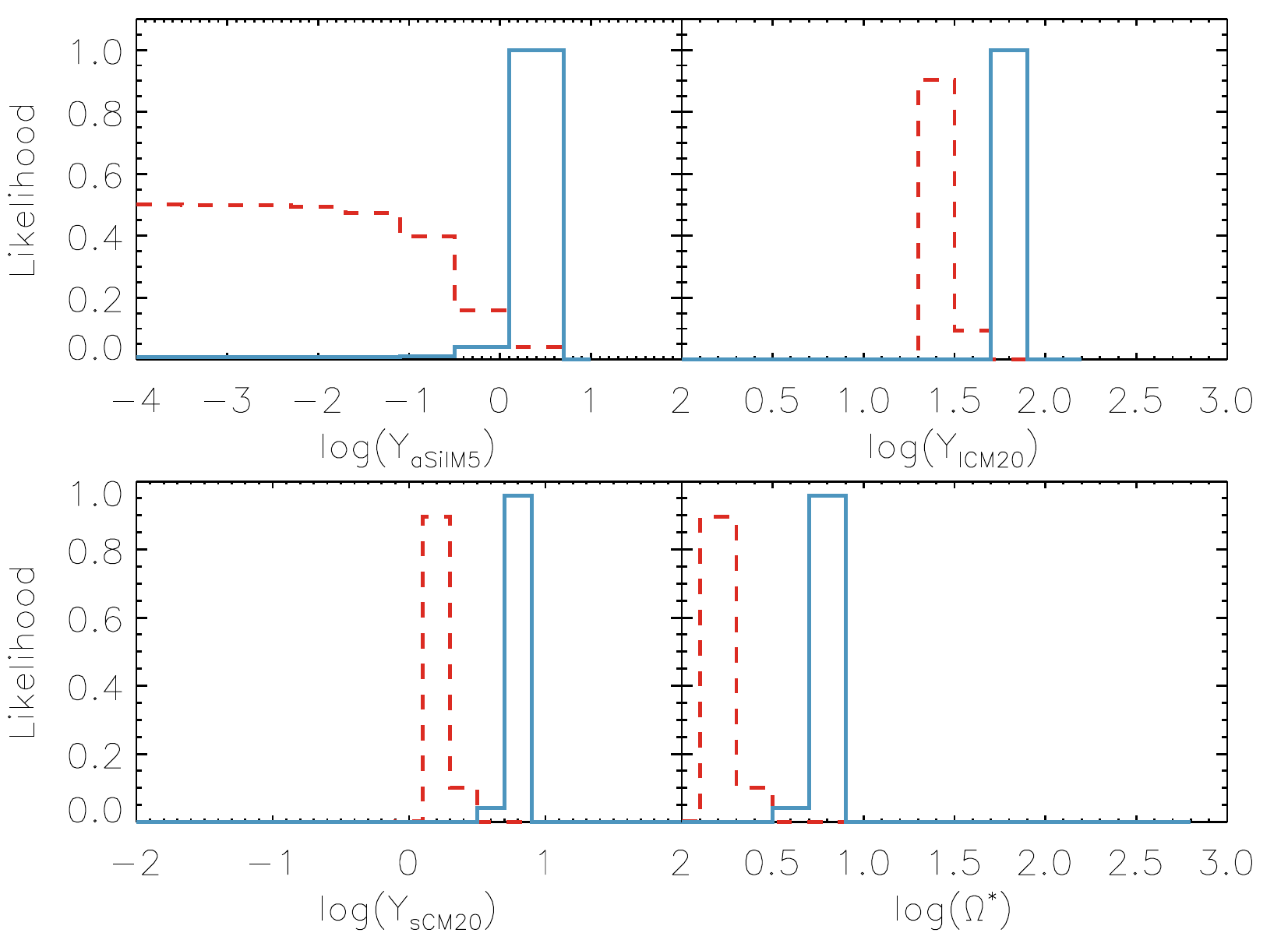}
    \caption{\small{Marginalized and normalized likelihoods of the $Y_\mathrm{aSilM5}$, $Y_\mathrm{lCM20}$, $Y_\mathrm{sCM20}$, and $\Omega^*$ parameters for two pixels in the SMC: a pixel showing an upper-limit in the $Y_\mathrm{aSilM5}$ fit (red-dashed line) and a pixel showing well-constrained fits (blue line).}}
    \label{FigLikelihoods}
\end{figure}

\begin{figure}
    \centering
        \includegraphics[width=0.45\textwidth]{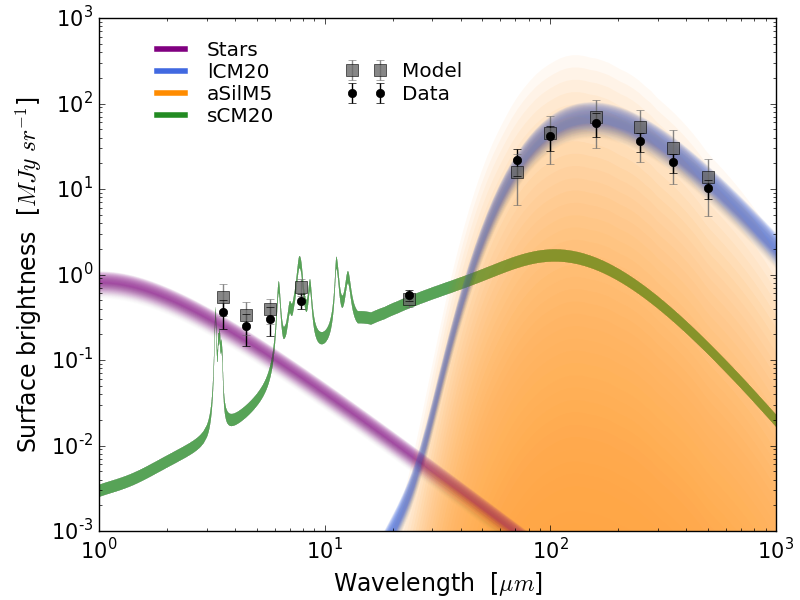}
        \includegraphics[width=0.45\textwidth]{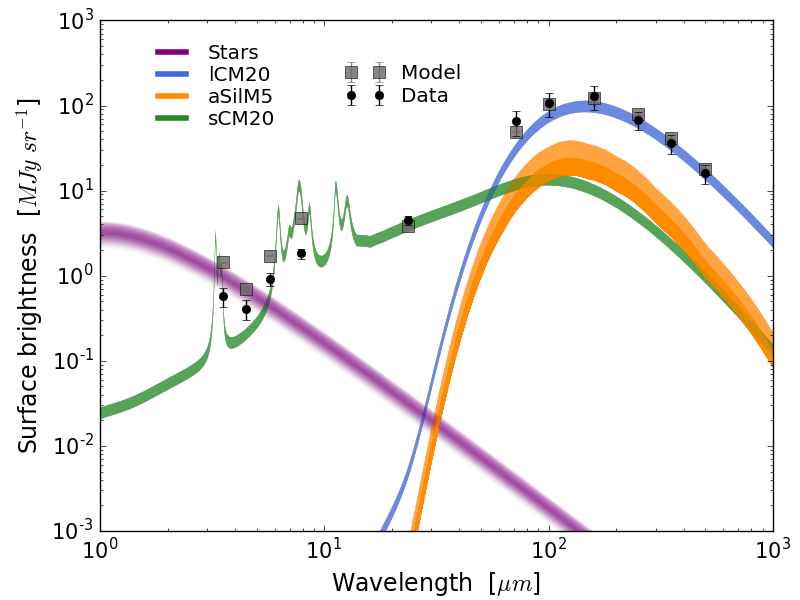}
    \caption{\small{Visualization of the unconstrained (top) and constrained (bottom) silicate fit. In the top panel, the different transparency surfaces show the possible range for the final silicate value, i.e. very broad and uncertain. On bottom panel, the same technique is used to draw a constrained fit.}}
    \label{FigSurfaceReal}
\end{figure}

In Figure \ref{FigLikelihoods}, we show likelihoods of the free parameters $Y_\mathrm{aSilM5}$, $Y_\mathrm{lCM20}$, $Y_\mathrm{sCM20}$, and $\Omega^*$ in two pixels: one that shows a good constraint on the amount of silicates (blue line), and one constraining  $Y_\mathrm{aSilM5}$ with an upper-limit only. The results in the two galaxies differ: in the LMC, $\sim{10\%}$ of the pixels show this kind of likelihood; in the SMC, more than $50\%$ do not show a fully-constrained fit.
In Figure \ref{FigSurfaceReal}, we display two representations of the SED fitting in the same pixels used for the likelihoods of Figure \ref{FigLikelihoods}. We used `realizations' of the likelihoods (as in \citet{Gordon14}, see Section \ref{SecModelComp}). The realizations are a weighted sample from the likelihood. 
The opacity of the color in Figure \ref{FigSurfaceReal} represents the probability of the value in the SED (the more opaque the color, the higher the probability). The top panel shows a very broad region with decreasing probability (i.e. increasing transparency), whereas the bottom panel depicts a constrained fit (i.e. opaque colors).
 
Using the upper limit, the silicate/carbon ratios for the variations on THEMIS (single-, 2-, multi-ISRFs and $\mathrm{\alpha_{sCM20}}$ free) ranges from $\sim 0.2 - 0.7$ in the SMC and $\sim 0.3 - 1.0$ in the LMC. The ratios vary from a model to another. There is only a slight evolution between the two galaxies, but this ratio considerably differs from that of the MW ($\sim 10$). In all cases, more than  97\% of the fit pixels exhibit a ratio well below the MW value.
It appears that the abundance of silicate component should not be kept constant in the MCs. 

In order to test the requirement of the silicate grains component in the fit, we perform fits using a single large grain species with THEMIS, either carbonaceous \emph{or} silicates, instead of allowing the two to vary. The large carbon species alone provides a good fit to the SMC IR peak: the residuals strictly follow the one obtained for THEMIS with a single ISRF and both, independent, grain components, and show no requirement for an additional silicate component. On the other hand, if we only allow a silicate population, we observe a very broad and multi-modal residual distribution. This is expected as the silicate emission is too narrow to fit the SMC IR peak between 100 and $500~\mathrm{\mu m}$. Unlike the SMC, the LMC SEDs requires both species to reproduce the data. A model without silicate grains follows the \emph{trend} of a ``complete'' model but does not match the data as well and a model without large carbon does not follow the observations.

We also perform a fit for which we tie the two populations, meaning that the silicate and the large carbon populations have to vary together the same way and keep the same initial ratio, that of the local ISM, given by \citet{Jones13} to be of $\sim 10$. In this case as well, the residuals are again large and bi-modal at long wavelengths. 

These results indicate that the silicate/carbon ratio in the SMC and LMC is not the same than in our Galaxy. 


\subsection{Dust Masses and Gas-to-Dust Ratios}
We compute total dust masses to assess if the models produce reasonable amounts of dust in each case.  We use multiple realizations of the likelihood in each pixel to estimate the total dust mass uncertainty. Contrary to the maximum likelihood or the expectation value, the realization samples the likelihood and therefore takes into account the contribution that the fitting noise, for each pixel, makes to the uncertainty in the total dust mass.

We create 70 maps from the realizations; where the sum of each of this map gives a total dust mass. The final total dust mass is the average of the realizations. Uncertainties on this value is given by the distribution of the total dust masses. We derive the total dust mass from pixels that are detected in 8 bands (see beginning of Section \ref{SecModelComp}) of the fit, at a level of at least 3$\sigma$ above the background noise. This corresponds to a surface of $\sim{2.1 \times 10^6}~\mathrm{pc}^2$ ($\sim 1.8~\mathrm{deg}^2$) in the SMC and $\sim{1.0 \times 10^7}~\mathrm{pc}^2$ ($\sim 14~\mathrm{deg}^2$) in the LMC. The dust masses are given in Table \ref{TabDustMasses}; they range from $\sim (2.9 - 8.9) \times 10^4~\mathrm{M}_\odot$ in the SMC and $\sim (3.7 - 4.2) \times 10^5~\mathrm{M}_\odot$ in the LMC, for THEMIS. We gather the results in the form of dust mass $\pm$ statistical uncertainty $\pm$ systematic uncertainty. The statistical uncertainty comes from the quality of the fits. It is very low due to the number of constraints we have. The systematic uncertainty comes from our understanding of the models and their limitations. More precisely, we mean the uncertainty on dust properties such as emissivity \citep[e.g.][]{WeingartnerDraine01, DraineLi07, Gordon14}, density, and the approach to build the optical and heating properties of dust grains (e.g. Mie theory, spherical grains). We also include degeneracies that come from our choices of ISRFs (e.g. a softer ISRF with more dust mass or stronger ISRF with less dust mass).

Given our constraints on the requirement for a pixel to be fit, there are a large number of `undetected' pixels. However, altogether, these regions may contribute significantly to the dust mass estimation. In order to take these pixels into account, we average their emission in each band to get an average SED. We fit this SED with the same models, and multiply by the surface area of all the undetected pixels to obtain a total dust mass. In the SMC, including the contribution from pixels below the detection threshold increases the dust masses given in Table \ref{TabDustMasses} from 50 to more than 100\%, doubling the mass in some cases. This is due to our very sparse pixel detection in this galaxy. The `undetected' area is about 10 times larger than the area covered by pixels detected. In the LMC, the `undetected area' is approximately the same size as the fitted area and accounts for $\sim 10-20 \%$ of the dust mass, \ion{H}{I} mass, and GDR for detected and undetected pixels are given in Table \ref{TabDustMasses}.

\renewcommand{\arraystretch}{1.17}
\begin{table*}[]
    \caption{Dust masses and GDR, in the SMC and the LMC.}
    \centering
        \begin{tabular}{ccccc}
        \hline
        \hline
        & \multicolumn{2}{c}{Pixels > 3$\sigma$ detection} & \multicolumn{2}{c}{Including pixels < 3$\sigma$ detection} \\
        \cline{2-5} 
        Model & M$_\mathrm{dust}$ [M$_\odot$] & GDR & M$_\mathrm{dust}$ [M$_\odot$] & GDR  \\
        \hline
        \multicolumn{5}{c}{SMC} \\
        \hline
        THEMIS single ISRF & $2.86 \pm 0.005 \pm 0.8 \times 10^4$ & $\sim 4100$ & $ 6.83 \pm 0.007 \pm 1.9 \times 10^4 $ & $\sim 1750$ \\
        THEMIS 2 ISRFs & $ 8.93 \pm 0.04 \pm 2.5 \times 10^4$ & $\sim 1300$ & $ 2.65 \pm 0.08 \pm 0.8 \times 10^5 $ & $\sim 500$ \\
        THEMIS multi-ISRFs & $7.68 \pm 0.02 \pm 2.3 \times 10^4$ & $\sim 1500$ & $ 1.20 \pm 0.07 \pm 0.3 \times 10^5 $ & $\sim 1000$ \\
        THEMIS $\mathrm{\alpha_{sCM20}}$ free & $6.25 \pm 0.01 \pm 1.7 \times 10^4$ & $\sim 1900$ & $ 1.01 \pm 0.009 \pm 0.3 \times 10^5 $ & $\sim 1200$ \\
        MC11 & $3.44 \pm 0.006 \pm 1.0 \times 10^5$ & $\sim 350$ & $ 3.71 \pm 0.006 \pm 1.1 \times 10^5 $ & $\sim 910$ \\
        \hline
        \multicolumn{5}{c}{LMC} \\
        \hline
        THEMIS single ISRF & $3.74 \pm 0.004 \pm 1.1 \times 10^5$ & $\sim 650$ & $ 4.51 \pm 0.008 \pm 1.3 \times 10^5 $ & $\sim 550$ \\
        THEMIS 2 ISRFs & $ 4.25 \pm 0.01 \pm 1.2 \times 10^5$ & $\sim 570$ & $4.89 \pm 0.04 \pm 1.4 \times 10^5 $ & $\sim 500$ \\
        THEMIS multi-ISRFs & $3.81 \pm 0.004 \pm 1.1 \times 10^5$ & $\sim 650$ & $ 4.73 \pm 0.005 \pm 1.4 \times 10^5 $ & $\sim 520$ \\
        THEMIS $\mathrm{\alpha_{sCM20}}$ free & $4.21 \pm 0.005 \pm 1.3 \times 10^5$ & $\sim 580$ & $ 4.88 \pm 0.008 \pm 1.4 \times 10^5 $ & $\sim 500$ \\
        MC11 & $2.05 \pm 0.001 \pm 0.6 \times 10^6$ & $\sim 120$ & $ 2.11 \pm 0.002 \pm 0.6 \times 10^6 $ & $\sim 170$ \\
        \hline
        \hline
        \end{tabular}
    \label{TabDustMasses}
    \begin{tablenotes}
        \item[1] Note: The total \ion{H}{} masses in the SMC and the LMC are $1.2 \times 10^8~\mathrm{M_\odot}$ and $3.3 \times 10^8~\mathrm{M_\odot}$ for pixels above the $3\sigma$ detection, and $2.4 \times 10^8~\mathrm{M_\odot}$ and $3.62 \times 10^8~\mathrm{M_\odot}$ when accounting for pixel below the $3\sigma$ detection.
  \end{tablenotes}
\end{table*}

\begin{figure}
    \centering
        \includegraphics[width=0.45\textwidth]{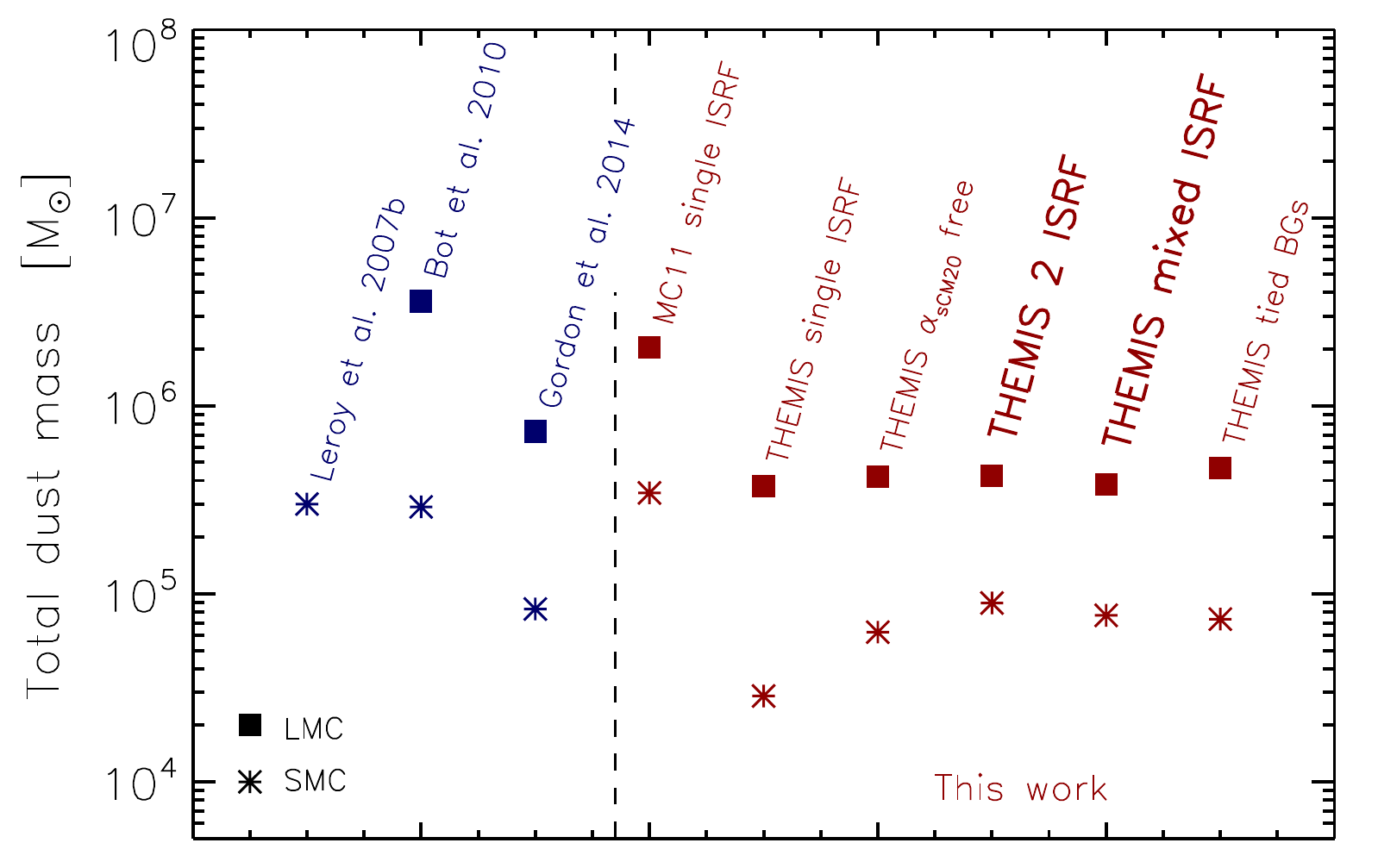}
    \caption{\small{Summary of the dust masses. Results from this work (right of the dashed line) are lower that previous studies. This likely comes from the low silicates abundances found in this paper.}}
    \label{dust_masses}
\end{figure}

In Figure \ref{dust_masses}, we gather some of dust masses found in the literature for the Magellanic Clouds and this work. 
The dust masses we derived in this study are smaller than those estimated by previous studies, especially for the simplest model. One interpretation of this 
difference lies in the carbon grains dominating our model fitting. The silicate emissivity is lower than that of large carbonaceous grains ($\sim 5~\mathrm{cm^2/g^{-1}}$ and $\sim 17~\mathrm{cm^2/g^{-1}}$ at 250~$\mathrm{\mu m}$, respectively). Therefore, for a same luminosity, if the fit uses only carbon grains, it requires less dust to produce the same flux than if it used both carbon and silicates. Because our best results indicate a very small contribution of silicates, this eventually leads to a lower dust mass. The environment, through the definition of the ISRF, seems to have a strong impact on the dust masses. Although it is hard to evaluate a quantitative difference with residuals (Section \ref{SecModelComp}), the final dust masses with a mixture of ISRFs is closer to the values found in other studies.

We perform a test fit to verify our assumption: in THEMIS, we tie the large grain populations (aSilM5 + lCM20) together, using a single-ISRF approach. In this process, we loose information regarding the independent distribution of the two types of grains, but it results in dust masses that are closer to those from \citet{Gordon14}, especially in the SMC, the LMC being only slightly affected by the change. This seems to confirm our assumption upon which the low dust mass we find come from our carbon grains-dominated fitting results.

The gas-to-dust ratio (GDR) estimation of a galaxy varies depending on the approach. Following \citet{RomanDuval14}, we determine GDRs using \ion{H}{I} measurements \citep{Stanimirovic00, Kim03}, and CO measurements \citep{Mizuno01} converted to H$_2$ mass estimations.
Our GDR estimations are thus really \emph{hydrogen}-to-dust ratio but for clarity, we keep the `GDR' notation.
We use the conversion coefficients $\mathrm{X_{CO}} = 4.7 \times 10^{20}$ cm$^{-2}$ (K km s$^{-1})^{-1}$ from \citet{Hughes10} for the LMC and $\mathrm{X_{CO}} = 6 \times 10^{21}$ cm$^{-2}$ (K km s$^{-1})^{-1}$  from \citet{Leroy07a} for the SMC. We report values of GDR in Table \ref{TabDustMasses}. As mentioned above, the dust masses with our model fitting are lower than the masses found by other works. This translates into higher gas-to-dust ratios. \citet{RomanDuval14} found GDR of $\sim{1200}$ for the SMC and $\sim 380$ for the LMC, using dust surface density maps from \citet{Gordon14}.
From Table \ref{TabDustMasses}, the GDR values derived from our favored fits (2-ISRFs and multi-ISRFs models) range from 1000 to 1200 in the SMC, and from 500 to 520 in the LMC. 
Using the model with tied large grains, we find GDRs lower than those found by previous studies ($\sim{700}$ in the SMC, $\sim{400}$ in the LMC), and the shape of the observed SED is not well reproduced.

Our GDRs show some variations (a factor of $\sim 2$ in the SMC between the higher and lower values). In order to assess a stronger constraint on the GDR, we compare these results to independent results given by depletion measurements or extinction.

Using the MW depletion patterns and the MCs abundances, one would expect GDRs of 540-1300 in the SMC, and 150-360 in the LMC. This assumes a similar dust composition and evolution between galaxies at different metallicities. We find values $\sim 2$ times higher than this. \citet{Tchernyshyov15} used UV spectroscopy to derive depletions in the MCs.  They found that scaling the MW abundances to lower metallicity, although approximately correct in the LMC, can lead to significantly different numbers in the SMC that those derived with depletions. From their results, they predict a range of GDRs: 480-2100 in the SMC and 190-565 in the LMC.  Our results 
fall within these limits. Their measurements were restricted to the diffuse neutral medium (DNM). Since we cover the diffuse to dense parts of the galaxy, one would expect more dust inferred from our fitting and thus, slightly lower GDRs.

Another way to predict GDR is to use extinction measurements. \citet{Gordon03} measured the dust extinction and \ion{H}{1} absorption column in the SMC and LMC, deriving N(\ion{H}{I})/A(V) values.  We determine the corresponding GDR expected from their results, using:
\begin{equation}
    \frac{1 / \mathrm{GDR}_{\mathrm{SMC}}}{1 / \mathrm{GDR}_{\mathrm{MW}}} = \frac{\mathrm{[N(\ion{H}{I})/A(V)]}_\mathrm{MW}}{\mathrm{[N(\ion{H}{I})/A(V)]}_\mathrm{SMC}}
\end{equation}
with $\mathrm{DGR}_{\mathrm{MW}}=1/150$. We used the averaged values in the SMC Bar, LMC and LMC2 (supershell) from their sample. We find reasonable values compared to their work.

Globally, our GDRs are in agreement with other studies that use different sets of measurements than IR emission. Our fits manage to reproduce the observed SEDs and fall within reasonable ranges for dust masses and GDR.
Previous studies have gathered GDRs estimations from numerous programs and estimated a trend between the metallicities of galaxies and their gas-to-dust mass ratios \citep[e.g.][]{Engelbracht08Err, Engelbracht08, Galametz11, Remy-Ruyer14}. We report our GDRs with the metallicity of the MCs \citep[12+log(O/H)$\sim 8.0$ in the SMC and $\sim 8.3$ in the LMC][]{RusselDopita92} and found that our values are in agreement with the trend. 

\section{Discussion}
\label{SecDiscussion}
\subsection{Grain formation/destruction}

The results from this study indicate that the silicate grains are not found in the same amounts in the LMC and the SMC with respect to carbon grains. The fits show that the silicate/carbon ratio is unlikely the same in the MW, LMC, and SMC. 

The lack of fully constrained fits of the silicate component suggests a deficit in silicate grain abundance, particularly in the SMC. 
This deficit could either be explained by less formation or by more destruction of silicate grains. \citet{Bocchio14} showed that silicate are less easily destroyed than carbon grains in supernovae (SNe) due to their higher material density.
This may therefore indicate that the higher abundance of carbon grains that we obtain is due to more efficient carbon dust formation rather than selective silicate destruction.
This could be consistent with the low metallicity of this galaxy. It is well established that carbon stars form more easily at low metallicity \citep[e.g.][]{Marigo08}. \citet{Nanni13} showed that such carbon stars are efficient producers of carbonaceous dust. With the carbon excess, O-type dust is unlikely to form due to the absence of M-type stars. Recent work by \citet{DellAgli15} investigated the evolution of AGB stars in the SMC using \textit{Spitzer} observations. Using color-color diagrams built from photometry and modelling, they identify C-rich and O-rich stars at various masses. They found discrepancies between their distribution in the LMC and SMC. 
The amount of O-rich AGB stars in these samples is lower in the SMC than that of in the LMC, which is $\sim{5 \%}$. This idea is coherent with depletion studies \citep[e.g.][]{Welty01, Tchernyshyov15}.

Yet, other studies have found constrains on the amount of silicates in the SMC. \citet{WeingartnerDraine01} constrained grain size distribution in the MW, LMC, and SMC, from elemental depletions and extinction curves. They adjust a functional form for each grain population (carbonaceous and silicate). In the case of the SMC, they reproduced the extinction curve toward AzV398 from \citet{GordonClayton98}, in the SMC-bar. Their result indicate a larger amount of silicate dust than carbon dust. Their result is therefore opposite to ours. However, we did not use the same approach, namely the nature of the observations, nor did we use the same dust models. In the next section we investigate the extinction curves in the MCs.

\subsection{Extinction curves}
Past programs measured extinction curves in the MCs, and have assessed discrepancies with the extinction curves in the MW (steeper far-UV slope, absence of $2175~\AA$ bump).
\citet{Gordon03} analyzed observed extinction curves in the MCs (5 in the SMC, and 19 in the LMC) and derived $R_V$ values. In their sample, most of the curves could not be reproduced using the relationship based on MW extinction curves. They found 4 curves in the LMC (Sk -69 280, Sk -66 19, Sk -68 23, and Sk -69 108) 
that show a MW-like extinction curve.

Our goal here is to verify if a fit of the dust emission in the MCs allows to reconstruct the observed extinction in the line of sight in our possession. We did \emph{not} try to directly fit the MCs extinction curves \emph{and} the corresponding SED in emission at the same time.
We extracted extinction curves at the same positions indexed in \citet{Gordon03} (4 in the SMC -- we did not fit the pixel corresponding AzV456 position, and all (19) in the LMC). Using the derived quantities for each grain species from our fits, we calculated extinction curves with the DustEM outputs. We only derived extinction curves for single-ISRF and the model where $\alpha_\mathrm{sCM20}$ is free. In the multi-ISRFs model, the dust composition does not change when we compute the mixture spectra, therefore each extinction curve is the same and we do not use this variation to infer conclusions.

\begin{figure}
    \centering
        \includegraphics[width=0.45\textwidth]{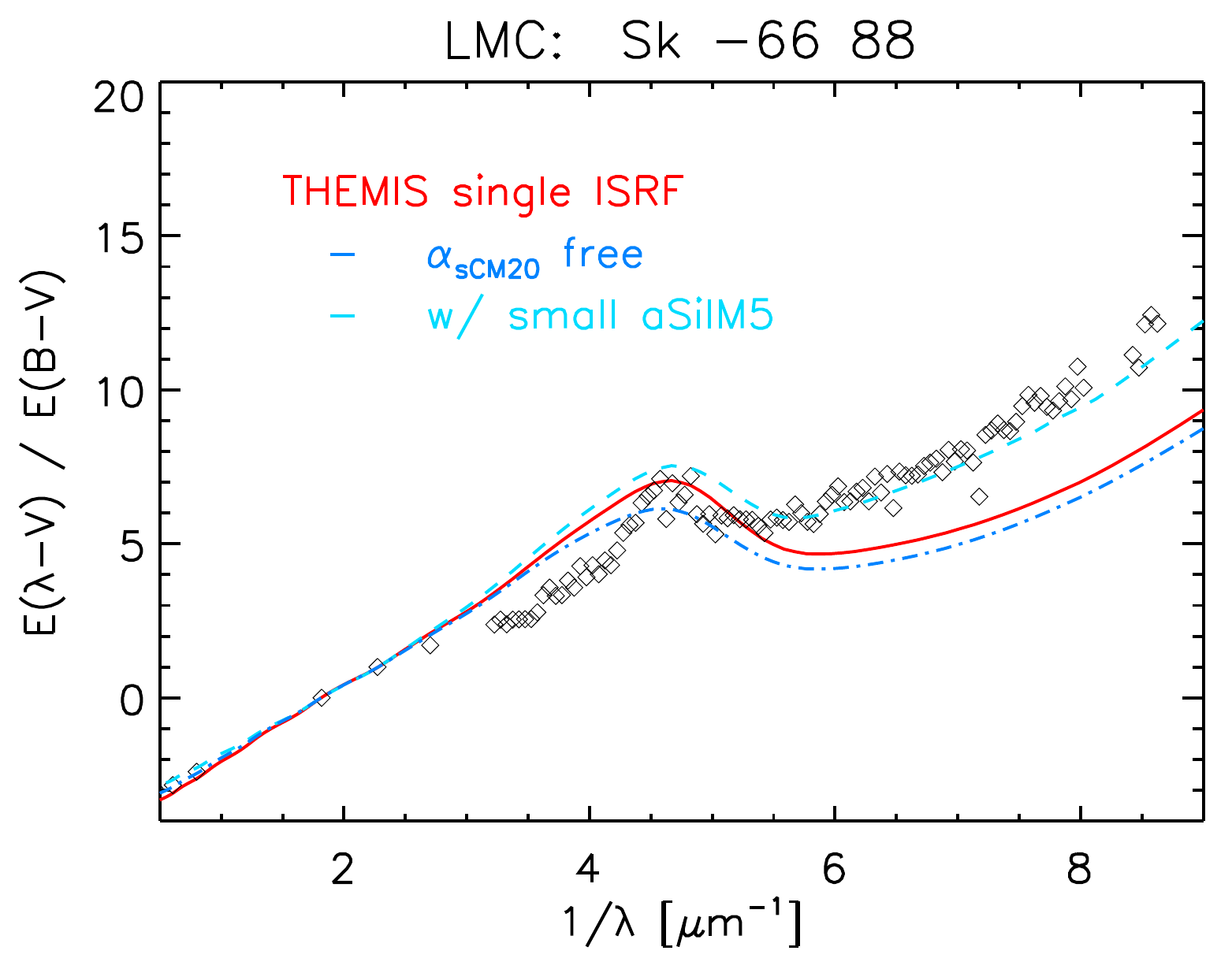}
        \includegraphics[width=0.45\textwidth]{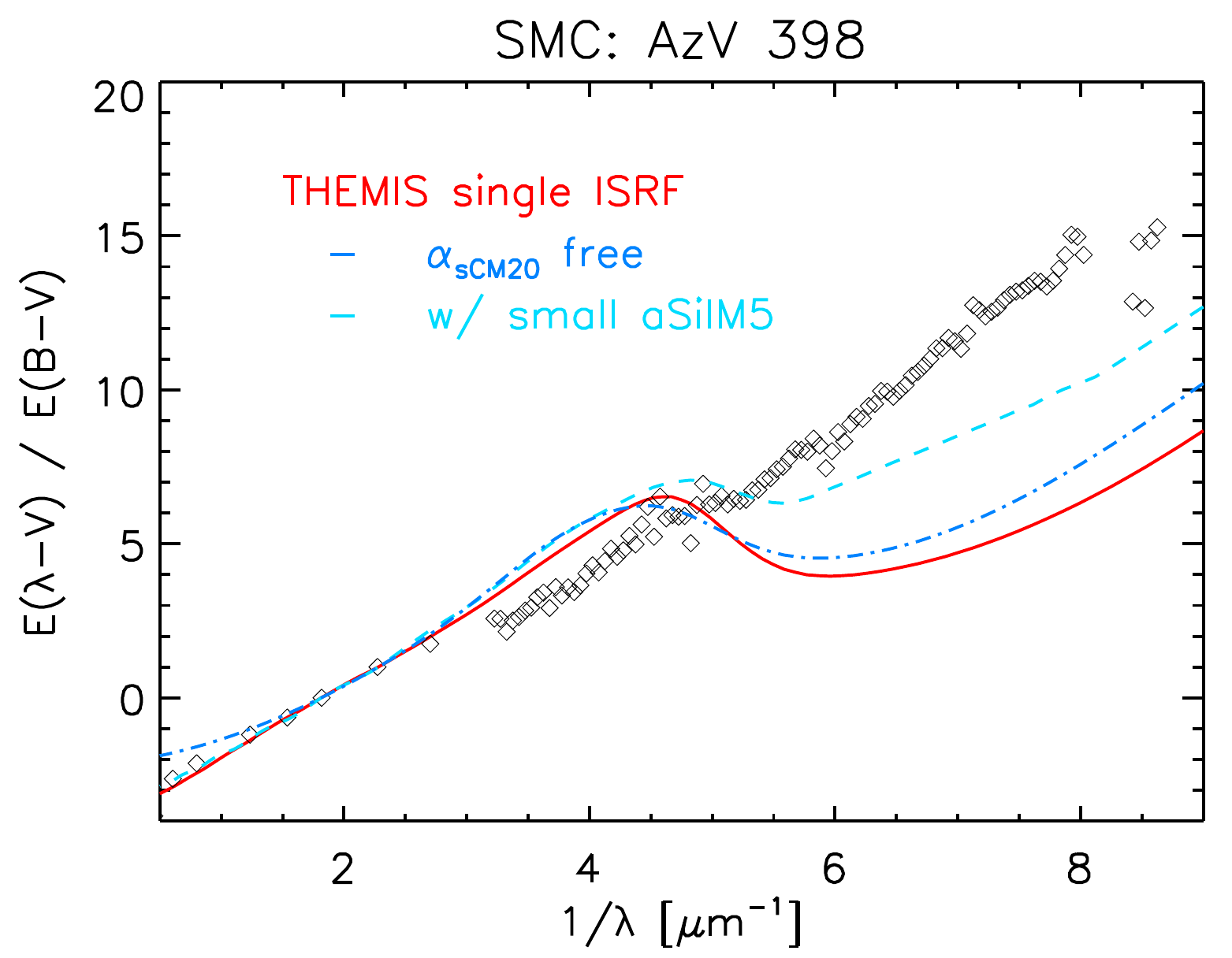}
    \caption{\small{Observed extinctions curves (black diamonds) in the LMC (\textit{top}: Sk -68 129) and the SMC (\textit{bottom}: AzV 398). We overplot the dust extinction derived from IR-emission fitting for a single-ISRF (red line) and a $\alpha_\mathrm{sCM20}$ free model (light blue-dash-dot line), and with smaller silicate grains (cyan-dashed line, see Section \ref{SecSmallaSil}).}}
    \label{FigExtCurves}
\end{figure}

In the LMC, we reproduce the extinction observations in the four lines of sight that showed a MW-like shape. In the SMC, none of the extinction curves can be reproduced using the dust population derived from IR emission fitting. This is also true for the rest of the LMC sample.
In Figure \ref{FigExtCurves}, we show two results of extinction curves derived from IR fitting. We plot the observed and modeled extinctions in AzV -68 129 (LMC), in the top panel. The $\alpha_\mathrm{sCM20}=5.4$ is very close to the default value in the single-ISRF model. Given the similar abundance values, the modeled extinctions are therefore comparable. In the SMC (bottom panel, AzV 398) as well, $\alpha_\mathrm{sCM20}=5.4$ and the result is close to that of a single-ISRF. In other lines of sight (e.g. AzV 214), a lower $\alpha_\mathrm{sCM20}$ (e.g. $\sim 3$) helps matching the near-IR part of the observed extinction ($1 \leqslant \lambda^{-1} \leqslant 3$). However, in both cases, the steep UV slope is not well fit at all.

When the shape of the extinction curve is different from that of the MW, IR emission does not predict the UV extinction. This could be due to the nature of dust grain models, that are based on a MW calibration. Or it could be due to a poor constraint on the small grain population by the IR emission, because the starlight and dust emission are mixed at those wavelengths. Either fitting the IR emission solely is not a good approach to derive a quantitative impact of the small grains on the extinction, or the small grain population needs to be split in order to derive various properties that do not affect emission and extinction the same way.
This result accounts for the differences one may find when fitting separately the dust emission and extinction. \citet{WeingartnerDraine01} used one extinction curve in the SMC (toward Azv398) to constrain a size distribution. They found a larger amount of silicate than carbon. In the same line of sight, our result from fitting the emission reproduces the observed extinction. However, allowing for a larger amount of silicate than carbonaceous grains, as suggested by \citet{WeingartnerDraine01} results, does not help to match the IR observations. 

\subsection{Other variations in dust models}
In our study, we investigated the change of model SED shape through variations in the ISRFs environments, and by allowing independent grain variations. We showed that such changes significantly increase the quality of the fits, especially in the SMC. Those variations mainly affect the $8-100~\mu$m range by steepening the $8/24~\mu$m slope and/or broadening the IR peak around $100~\mu$m. 
Other studies can provide more suggestions to change the composition of dust models.

\subsubsection{Change in carbon size distribution}
\citet{Kohler15} studied the dust properties evolution from diffuse to dense regions. For example, they showed that grains with an additional mantle have different properties that can lead to a steepening of the FIR slope and a lower temperature. They also investigated the influence of forming aggregates in dense media. In that case as well, the dust properties vary significantly.
Such approaches could be helpful to fit the MCs dust emission. In Figures \ref{FigLMC_AJMCs} --- \ref{FigLMC_AJsa}, we can see THEMIS is slightly above the observations in the FIR (at 250, 350, 500~$\mu m$). A steepening of the spectral index may suggest that the model would better match the data at these wavelengths.

\citet{Ysard15} also investigated the variations of dust properties observed with \textit{Planck}-HFI. In their study, they investigated the impact of varying the carbon abundance, while keeping the silicate abundance constant. They found that this variation could help reproduce the observation and account for the dust variations. They showed that changing the size distribution (by changing the aromatic-mantle thickness or the size distribution function) participate in the dust variations. This provides additional evidence that a single model with fixed size distribution is not appropriate for fitting observations on a galaxy scale.

\subsubsection{Allowing smaller silicate grains}
\label{SecSmallaSil}
\citet{Bocchio14} computed size distribution, emission and extinction curves for carbonaceous and silicate grains from THEMIS in environments where dust is destroyed/sputtered by shocks with $v \sim 50 - 200~\mathrm{km/s}$. For sufficiently high shock velocities carbon grains are highly destroyed, whilst silicates are fragmented into smaller grains due to their collisions with small carbon grains. Using their silicate grain size distribution leads to a steepening of the far-UV extinction. Looking at the peculiar shape of the SMC extinction, we find this approach interesting. Once again, we want to know if fitting the dust emission can yield to a good estimation of the dust extinction.

Allowing for smaller silicate grains helps to match the data in the IRAC bands, in the SMC. In the LMC, on the other hand, the residuals are not affected significantly, and the fits are not improved.  However, in both galaxies, we can notice a change in the extinction shape. The far-UV slope is closer to the observations. In Figure \ref{FigExtCurves}, the cyan-dashed lines are those derived for a fit of the emission with smaller silicate grains. In the SMC, twy lines of sight are significantly improved by the change in the silicate size distribution. Our results still exhibit a small bump around $2175~\AA$, because we allow the small carbonaceous grains to vary. In the LMC, some lines of sight are greatly affected and the extinction can be matched with smaller silicates, e.g. top panel of Figure \ref{FigExtCurves}.

We only applied the new size distribution in a single-ISRF environment, so we can test the resulting extinction with DustEM. In terms of dust masses, the new fit leads to $\sim 4.4 \times 10^4~\mathrm{M_\odot}$ in the SMC, and $\sim 2.0 \times 10^5~\mathrm{M_\odot}$ in the LMC, respectively higher and lower than a single ISRF environment, without the change in size distribution.

\subsubsection{On the recalibration}
In our study, we used a different reference SED to rigorously compare models after they were recalibrated on the same Galactic values. However, it should be noticed that the models are not defined as such. In THEMIS, the GDR is set to $\sim 134$ \citep[][]{Ysard15}.
Without recalibration we would obtain a dust mass of $\sim (1.9-5.6) \times 10^4~\mathrm{M_\odot}$ in the SMC, and $\sim (2.3-2.7) \times 10^5~\mathrm{M_\odot}$ in the LMC, for the different variations of environment. THEMIS mass distribution for the grain populations is different than other dust models \citep[e.g.][]{DraineLi07}. For example, the silicate (pyroxene and olivine type) grains have a lower specific mass density. Therefore, the model needs less silicate mass. We can also notice that the carbon mass is mostly found in small carbonaceous grains.

In order to get a more accurate mass estimation, one possible path of investigation is to use the different versions of the model to fit the various media of a galaxy, namely dense or diffuse. In THEMIS, dust in the transition from the diffuse ISM towards dense molecular clouds is described with aSilM5 and lCM20 grains coated with an additional H-rich carbon mantle. Inside dense molecular clouds, further evolution is assumed and THEMIS dust consists in aggregates (with or without ice mantles).



\section{Conclusions}
We fit the \textit{Spitzer} SAGE and \textit{Herschel} HERITAGE observations of the Magellanic Clouds at $\sim10$ pc, in 11 bands from 3.6 to 500~$\mu m$. We use two physical dust grain models: \citet{Compiegne11} and THEMIS \citep[][]{Jones13,Kohler15} to model dust emission in the IR.

On global evidence, we find that the \citet{Compiegne11} should not be used in this context as it suffers from strong discrepancies with respect to the observations (e.g. big grain steep emissivity in the FIR).
Fitting THEMIS on the observations gives better residuals, especially in the SMC. THEMIS leaves a small deficit in the residuals in the FIR, i.e. is too low compared to the observations, in opposition to what has been identified in previous studies as an excess at 500~$\mu m$. 
We find that using more than a single ISRF greatly improves the quality of the fit. More generally, a change in the shape of the model SED will help to get better residuals, either by using more than a single-ISRF environment, or by changing the dust grain size distribution, with respect to the one calibrated on the diffuse ISM in the solar neighborhood. 
Parameters maps depict very model-dependent spatial variations. The approach chosen for the dust environment (ISRF) strongly affects the quality and results of the fits.

Using THEMIS dust model, we find than the silicate abundance is estimated only as an upper-limit $Y_\mathrm{{aSilM5}} \sim 10^{0.4} M{^\mathrm{aSilM5}}{_\odot} / M_\mathrm{H}$, while the large carbonaceous grain emission is constrained with well defined peaked likelihood distributions. The silicate/carbon ratio implied by the fits indicate an evolution between the MW, and the MCs. This ratio is $\sim{10}$ in the MW, but is not the same nor constant throughout the MCs ($\leqslant 1$ in the SMC and LMC). Tests forcing a MW-like silicate/carbon ratio leads to very broad residuals and poor fitting, confirming that this ratio should not be kept constant for these galaxies.

The dust masses derived in the LMC from our fitting are lower than those derived by other studies by a factor lower than 2, but remain close given our uncertainties (of $\sim 30\%$ total). In the SMC, our values are in agreement with literature \citep[e.g.][]{Gordon14} but suffer from large uncertainties.
The numerous pixels with the low upper-limit are mostly responsible for the slightly lower dust masses (especially in the SMC). 

We used our dust emission results to create \emph{modeled} extinction curves. We find that fitting only the emission cannot give results to apply directly to match the measured dust extinction in the MCs.
These tests showed that a change in the estimated grain size distributions (based on MW measurements), would be needed to (more) accurately match the MCs extinction, from an emission fitting (e.g. different silicate grain distribution, namely smaller).

Further work will use additional dust grain models to be compared with \citep[e.g.][THEMIS with aggregates]{DraineLi07}, while the goal should remain the same, i.e. compared dust emission/extinction results from various dust models using a strictly identical fitting technique. In order to fully interpret these data, a more detailed phase-specific approach is needed (but is beyond the scope of this paper). Radiative transfer technique should also be used to understand the systematics of the assumptions made when developing a dust model in a given environment (e.g. `single-U' or `multi-U').

\begin{acknowledgements}
We would like to thank the referee S. Bianchi for a careful and thorough reading that caught mistakes and phrasing confusions, and considerably helped improving the paper.
\end{acknowledgements}

\end{document}